\documentclass[twocolumn]{aastex63}

\usepackage{CJK}

\shorttitle{HR 8799 Atmospheric Monitoring}
\shortauthors{Wang et al.}

\begin{document}
\begin{CJK*}{UTF8}{gbsn}

\title{Atmospheric Monitoring and Precise Spectroscopy of the HR 8799 Planets with SCExAO/CHARIS\footnote{Based on data collected at Subaru Telescope, which is operated by the National Astronomical Observatory of Japan.}}
\correspondingauthor{Jason Wang}
\email{jwang4@caltech.edu}

\author[0000-0003-0774-6502]{Jason J. Wang (王劲飞)}
\altaffiliation{51 Pegasi b Fellow}
\affiliation{Department of Astronomy, California Institute of Technology, Pasadena, CA 91125, USA}
\affiliation{Center for Interdisciplinary Exploration and Research in Astrophysics (CIERA) and Department of Physics and Astronomy, Northwestern University, Evanston, IL 60208, USA}

\author[0000-0002-8518-9601]{Peter Gao}
\affiliation{Earth \& Planets Laboratory, Carnegie Institution for Science, 5241 Broad Branch Rd NW, Washington, DC 20015, USA}

\author[0000-0001-6305-7272]{Jeffrey Chilcote}
\affiliation{Department of Physics and Astronomy, University of Notre Dame, 225 Nieuwland Science Hall, Notre Dame, IN, 46556, USA}

\author[0000-0002-3047-1845]{Julien Lozi}
\affiliation{Subaru Telescope, NAOJ, 650 North A{'o}hoku Place, Hilo, HI 96720, USA}

\author[0000-0002-1097-9908]{Olivier Guyon}
\affiliation{Subaru Telescope, NAOJ, 650 North A{'o}hoku Place, Hilo, HI 96720, USA}
\affiliation{Steward Observatory, University of Arizona, 933 N Cherry Avenue, Tucson, AZ 85719, USA}
\affiliation{Astrobiology Center of NINS, 2-21-1 Osawa, Mitaka, Tokyo 181-8588, Japan}

\author[0000-0002-4164-4182]{Christian Marois}
\affiliation{National Research Council of Canada Herzberg, 5071 West Saanich Rd, Victoria, BC, V9E 2E7, Canada}
\affiliation{University of Victoria, 3800 Finnerty Rd, Victoria, BC, V8P 5C2, Canada}

\author[0000-0002-4918-0247]{Robert J. De Rosa}
\affiliation{European Southern Observatory, Alonso de C\'{o}rdova 3107, Vitacura, Santiago, Chile}

\author[0000-0003-2806-1254]{Ananya Sahoo}
\affiliation{Space Telescope Science Institute, 3700 San Martin Drive, Baltimore, MD 21218, USA}
\affiliation{Subaru Telescope, NAOJ, 650 North A{'o}hoku Place, Hilo, HI 96720, USA}

\author[0000-0001-5978-3247]{Tyler D. Groff}
\affiliation{NASA-Goddard Space Flight Center, Greenbelt, MD, USA}

\author[0000-0003-4018-2569]{Sebastien Vievard}
\affiliation{Subaru Telescope, NAOJ, 650 North A{'o}hoku Place, Hilo, HI 96720, USA}

\author[0000-0001-5213-6207]{Nemanja Jovanovic}
\affiliation{Department of Astronomy, California Institute of Technology, Pasadena, CA 91125, USA}

\author[0000-0002-7162-8036]{Alexandra Z. Greenbaum}
\affiliation{IPAC, MC 100-22, Caltech, 1200 E. California Blvd., Pasadena, CA 91125, USA}

\author[0000-0003-1212-7538]{Bruce Macintosh}
\affiliation{Kavli Institute for Particle Astrophysics and Cosmology, Stanford University, Stanford, CA 94305, USA}

\begin{abstract}

The atmospheres of gas giant planets are thought to be inhomogeneous due to weather and patchy clouds. We present two full nights of coronagraphic observations of the HR 8799 planets using the CHARIS integral field spectrograph behind the SCExAO adaptive optics system on the Subaru Telescope to search for spectrophomometric variability. We did not detect significant variability signals, but placed the lowest variability upper limits for HR 8799 c and d. Based on injection-recovery tests, we expected to have a 50\% chance to detect signals down to 10\% $H$-band photometric variability for HR 8799 c and down to 30\% $H$-band variability for HR 8799 d. We also investigated spectral variability and expected a 50\% chance to recovery 20\% variability in the $H/K$ flux ratio for HR 8799 c. We combined all the data from the two nights to obtain some of the most precise spectra obtained for HR 8799 c, d, and e. Using a grid of cloudy radiative-convective-thermochemical equilibrium models, we found all three planets prefer supersolar metallicity with effective temperatures of $\sim$1100 K. However, our high signal-to-noise spectra show that HR 8799 d has a distinct spectrum from HR 8799 c, possibly preferring more vertically extended and uniform clouds and indicating that the planets are not identical. 
\end{abstract}

\keywords{Exoplanet atmospheres ({487}), Exoplanet atmospheric variability ({2020}), Coronagraphic imaging ({313}), Spectrophotometry ({1556})}

\section{Introduction} \label{sec:intro}

Atmospheric spectroscopy provides a key window into understanding the nature of exoplanets. Through the measurement of molecular absorption features, we can learn about the composition of planets to uncover their nature and their formation history. However, molecular absorption is not the only feature in planetary spectra. Clouds are another central aspect of planetary atmospheres. Clouds trace the weather on other planets, and the detailed monitoring of their clouds can help us understand exoplanet atmospheric dynamics \citep{Tan2017}. They can also alter molecular absorption feature strengths, and affect our understanding of the composition of other planets \citep{Burningham2017, Molliere2020}. 

The atmospheres of directly imaged planets are generally similar to higher mass brown dwarfs of the same effective temperature \citep[e.g.,][]{Chilcote2017}.Planets of the L and T spectral type have redder spectra in the near-infrared than the higher-mass brown dwarfs, which could be due to having thicker clouds \citep[e.g.,][]{DeRosa2016}. For the four super-Jupiters orbiting HR 8799 \citep{Marois2008, Marois2010}, we have accumulated a diverse set of photometry and spectra that point to a complex cloud model \citep{Bowler2010, Barman2011, Konopacky2013, Currie2014, Ingraham2014, Skemer2014, Zurlo2016, Bonnefoy2016, Greenbaum2018, Molliere2020, Ruffio2021, Wang2020, Wang2021}. In particular, the temperature of the four planets lie around the transition between the L and T spectral types, where partly cloudy atmospheres are thought to induce the enhanced photometric variability ($>2$\% in $J$ band) we see in free-floating substellar objects at this temperature range \citep{Radigan2014}. 

If patchy clouds exist in these planetary atmospheres and the planets are not viewed pole-on, we expect photometric variability in the light curves of these planets as different patches of clouds rotate into and out of view \cite{Vos2017}, although exoplanet weather could induce some variability signal even for planets viewed pole-on \cite{Tan2021}. Rotational variability has been seen in the light curves of wide separation and free-floating planetary mass objects with nearly the same mass and atmospheric properties as the HR 8799 planets \citep{Biller2018,Bowler2020,Zhou2020}. There has also been indirect evidence for this in the HR 8799 planets as patchy cloud models have fit 3 to 5 $\mu$m photometry of the HR 8799 planets better \citep{Skemer2014}. This has motivated monitoring programs for the HR 8799 planets, but no variability has been detected \citep{Apai2016,Biller2021}. One issue is the difficulty in performing precise photometry in these high contrast imaging datasets, stemming from effectively removing the glare of the host star as well as obtaining precise photometric calibration with the host star. Due to this, the current photometric precision obtained on the HR 8799 planets is an order of magnitude away from what has been obtained for free-floating planetary mass objects \citep{Biller2018,Biller2021}.

In this paper, we present the results of monitoring the HR 8799 planets from Maunakea, where the system is visible for the entirety of our two full nights, using the Subaru/CHARIS integral field spectrograph. In Sections \ref{sec:obs} and \ref{sec:data}, we describe the observations and data reduction respectively. We discuss the stability of the spectrophotometric calibration in Section \ref{sec:satspots}. We do not see any significant periodic variability signatures and place limits on the photometric and spectral variability of the planets in Section \ref{sec:var}. However, we were able to combine all the spectra we obtained to precisely measure the near-infrared spectra of the inner three planets (HR 8799 c, d, and e) and fit atmospheric models to constrain the properties of these three planets in Section \ref{sec:atm-fits}. 

\section{Observations} \label{sec:obs}

\subsection{HR 8799}
HR 8799 was observed at the Subaru telescope on two consecutive nights for nearly the entire time using the CHARIS integral field spectrograph \citep{Groff2015,Groff2017} behind the SCExAO high order adaptive optics system with its Lyot coronagraph \citep{Jovanovic2015_scexao}. We generated ``satellite spots" with the deformable mirror to create calibration point sources with attenuated copies of the stellar spectrum in the image since the star is behind the coronagraph. We quickly modulated the deformable mirror to make the spots incoherent with the speckle field \citep{Jovanovic2015}. We obtained 20~s exposures using the CHARIS low-resolution mode that obtained R$\sim$20 spectra from $J$ through $K$ band. On 2018 September 1 (night 1), we obtained 1,201 usable frames of data from 7:03 to 15:34 UT, resulting in a total on-sky integration time of 24,805~s (6.89~hours). On 2018 September 2 (night 2), we obtained 1,253 usable frames of data from 6:11 to 15:45 UT after discarding 89 frames with unusual detector noise, resulting in a total on-sky integration time of 25,879~s (7.19~hours). Altogether, we obtained 14.08 hours of integration time on HR 8799 over the course of the two nights. We note that a portion of this data was used in \citet{Wang2020} to study HR 8799 c specifically, and here we will use this data to study all three planets in our field of view. 

\subsection{HD 187003}
On each night, we observed the stellar binary HD 187003 beforehand to calibrate the brightness and stability of the satellite spots generated by the deformable mirror. HD 187003 has a secondary located at $\sim$0.6~arcsec away from the primary with a $K$-band flux ratio of $\sim3 \times 10^{-2}$. This stellar companion is favorable for studying the stability of the satellite spots, since it is inside the field of view and not so bright that it saturates the image. 

On 2018 September 1, we obtained 30 frames with 15~s integration times in the same low-resolution mode with the Lyot coronagraph as the HR 8799 data. On 2018 September 2, we obtained a longer 87 frame sequence with the same configuration. Furthermore, on 2018 September 2, we took three 200~s images where the primary star was not behind the coronagraph, but the whole field of view was attenuated by a neutral density filter to prevent saturation. These images were used to directly measure the flux ratio of the stellar binary at each wavelength. With the binary flux ratio known, we can then calibrate the satellite spot flux ratio in each exposure where the binary is occulted.

\section{HR 8799 Data Reduction}\label{sec:data}

\subsection{Spectral Datacube Processing}

\begin{figure*}
    \centering
    \includegraphics[width=\textwidth]{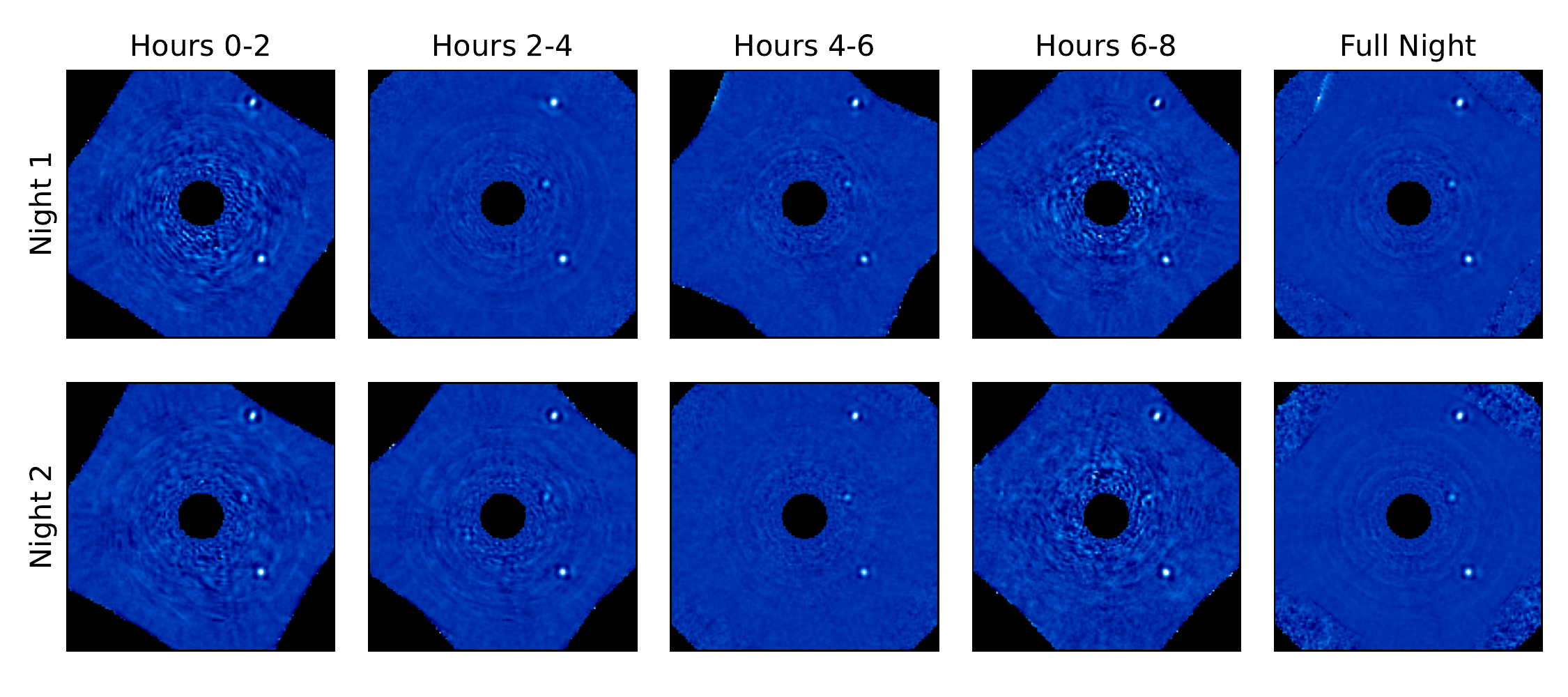}
    \caption{Reduced images of HR 8799 in two-hour chunks and full eight-hour temporally-collapsed images. The images were combined using a weighted mean where each wavelength slice of each exposure is weighted by the noise at the separation of HR 8799 e in that image. All the images are in linear scale, scaled to the flux of HR 8799 c, but not corrected for signal loss due to post-processing (i.e., algorithm throughput).  }
    \label{fig:hr8799_gallery}
\end{figure*}

First, we will discuss the processing steps to generate and calibrate the 3D spectral datacubes. The initial steps were performed using the CHARIS data reduction pipeline \citep{Brandt2017}. The raw 2D detector frames consist of an array of microspectra, with each corresponding to one pixel of the final 3D datacube. The 2D images were dark subtracted using dark frames taken at the end of each night. Using a narrow-band tuneable filter, point spread functions (PSFs) of the microspectra at each wavelength were taken at the end of each night and compiled into a wavelength solution for each spectrum. The microspectra were then extracted using the algorithm in \citet{Brandt2017} to create 3D spectral datacubes. The datacubes have 22 spectral channels, with each channel corresponding to an image of HR 8799 at a specific wavelength. Each datacube was then flat fielded using a flat field cube where all of the microspectra were illuminated by a flat lamp. Note that the whole 2D detector was not illuminated by the flat field, so this flat field is a combination of detector response, lenslet response, and cube extraction effects. 

Following, we located the position of the four satellite spots in each wavelength channel of each datacube to determine the position of the occulted star and to extract off-axis PSFs. We fit each of the four satellite spots in each wavelength channel of each cube with a 2D Airy function. We fit the position, peak flux, and width of the Airy function using Nelder-Mead minimization. We averaged the position of the four satellite spots to calculate the location of the star behind the occulting mask as a function of wavelength. We also used the position of each spot to extract out a 15x15 pixel cutout of each satellite spot. We interpolated each spot so that it is centered on the middle pixel of the cutout, and subtracted the background in the cutout by taking the average background value between 9 and 12 pixels away from the center of the satellite spot. We combined the cutouts of the four spots to create a single satellite spot PSF for each frame.  

\subsection{Stellar PSF Subtraction}
In order to look for temporal variability, we broke up the observing sequences into one- and two-hour chunks for the analysis. We chose these two binning strategies because two-hour chunks gave us better parallactic angle rotation for ADI, but one hour chunks gave us better temporal resolution. Regardless, for each chunk, we performed stellar PSF subtraction using only data from that chunk to remove the glare of the host star. We used the open-source Python package \texttt{pyKLIP} \citep{Wang2015} to model and subtract off the glare of the star using principal component analysis \citep[PCA,][]{Soummer2012}. We employed both angular differential imaging (ADI) and spectral differential imaging (SDI) to select images of the system taken at other times and wavelengths to create reference images of the stellar PSF \citep{Marois2000, Sparks2002, Liu2004, Marois2006}. The PCA modes were constructed using the 200 most correlated reference PSFs from these reference images. For each planet, we varied the minimum number of pixels the planet has to move due to ADI and SDI for that frame to be included in the reference images and the number of PCA modes used. We create a fiducial image using 50 PCA modes and used reference images where the planets moved at least 1 pixel due to ADI and SDI. The two-hour-chunk reduction are plotted in Figure \ref{fig:hr8799_gallery}. HR 8799 c and d are clearly detected in all images, but HR 8799 e is marginally detected in the first and last two hour chunk due to the limited parallactic angle rotation for ADI. This is consistent with previous findings that ADI is more effective than SDI for SCExAO/CHARIS PSF Subtraction \citep{Gerard2019}.

\subsection{Spectral-photometric Extraction}\label{sec:spectral_extraction}

\begin{figure*}
    \centering
    \includegraphics[width=\textwidth]{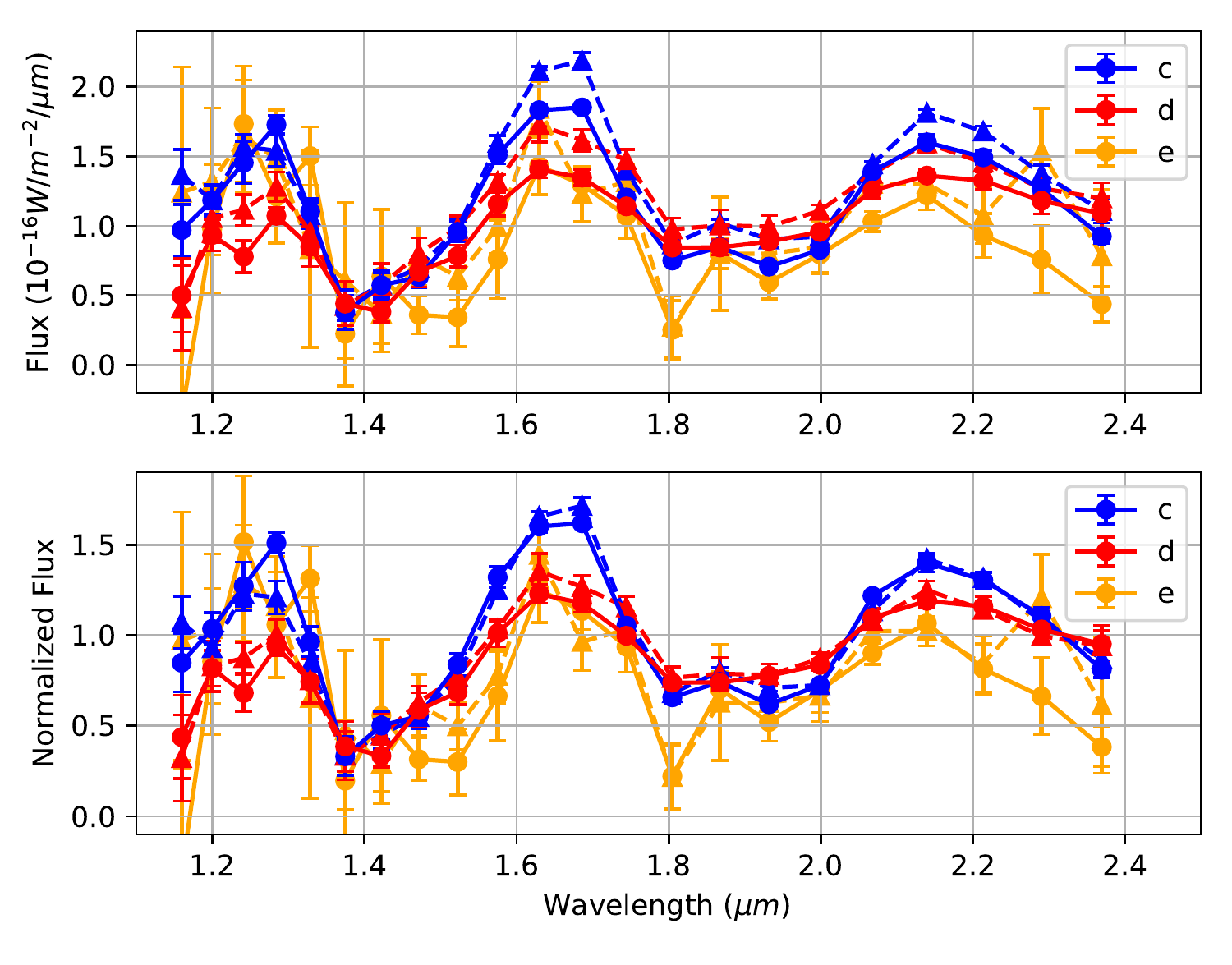}
    \caption{Top: Spectra of HR 8799 c, d, and e resulting from summing 8 hours of data from each night. Night 1 spectra are plotted with solid lines and night 2 spectra are plotted with dashed lines. The error bars only reflect the measurement error of the spectrum and does not include systematic errors due to spectro-photometric calibration which are discussed in Section \ref{sec:satspots}. Bottom: The same spectra normalized to the average $K$-band flux of HR 8799 c in each night. The calibrated spectra of the three planets in the top panel are available as the Data behind the Figure. }
    \label{fig:hr8799_spec_fullnight}
\end{figure*}

Next, we measured the flux of each planet at each wavelength in the CHARIS observations. Due to the known effects of overfitting caused by PCA, we used the forward modeling framework presented in \citet{Pueyo2016} and \citet{Greenbaum2018} and implemented in \texttt{pyKLIP} to obtain the spectrum of each planet in each chunk of data. The forward modelling utilizes the fact we know the instrumental PSF (derived from the satellite spot PSFs), how the planet moves in the images due to ADI and SDI, and a linear approximation for how the planet perturbs the PCA fitting to self-consistently recover the ratio of the planet's flux at each wavelength to the instrumental PSF flux (i.e., satellite spot flux). As there is a concern that our linear approximation of each planet's effect on the stellar PSF subtraction may not be valid for precision variability monitoring of bright planets \cite{Pueyo2016}, we injected eight simulated planets at the same separation as each of the three real planets but at different position angles (we only injected one simulated planet at a time to avoid cross talk). Each simulated planet was injected into the data with the spectrum of the real planet we had previously extracted from the data to accurately measure any biases from spectral extraction. We then extracted the spectrum of each simulated planet using the same method. 

For each real planet at each wavelength in each chunk of data, we computed a bias factor that was the average ratio of the measured simulated planet flux with the input simulated planet flux. We found biases between the injected and measured fluxes of up to 10\%. We used the measured bias factors to correct our HR 8799 planet spectra. We computed uncertainties for the flux of each real planet at each wavelength in each one- and two-hour chunk by computing the standard deviation of the measured simulated planet fluxes and correcting for the bias term. To convert the flux measurement from a ratio of planet flux to satellite spot flux to the planet's flux in physical units, we use the satellite spot flux ratios for each night calculated in Section \ref{sec:satspot-phot} to derive a planet to star flux ratio and a 7330~K PHOENIX stellar model and the star's $K$-band magnitude of 5.24 \citep{Cutri2003} to derive the spectrum of each planet in $\textrm{W}/\textrm{m}^2/\mu\textrm{m}$. The 7330~K PHOENIX spectrum agrees well with the effective temperature of the star found in other recent work \citep{Wang2020}. We applied this process to both the one-hour and two-hour chunking of the data. 

To obtain the highest SNR spectrum of each planet from each night, we then combined the two-hour-chunk spectra in time using a weighted mean where the weights are the inverse square of the uncertainties. This is nearly identical to running the data analysis on an entire night of data, but saves on computation time. We derived the errors on these stacked spectra using the formal uncertainty propagation for a weighted mean. We plot these spectra for all three planets in Figure \ref{fig:hr8799_spec_fullnight}. For HR 8799 c in $H$- and $K$-band, we reach signal-to-noise ratios (SNRs) $> 50$ per spectral channel. We note that the plotted errors consist only of measurement uncertainty and does not include systematic errors in the spectro-photometric calibration that we will explore in Section \ref{sec:satspots}. We can see in Figure \ref{fig:hr8799_spec_fullnight} that the spectra from night 2 appear to be systematically brighter than the spectra from night 1 if we take the photometric calibration at face value. Interestingly, if we normalize each night's spectra to the $K$-band flux of HR 8799 c that night, then the three spectra have better agreement in Figure \ref{fig:hr8799_spec_fullnight}, indicating that the first order correction is a constant scale factor across all wavelengths between nights. This could be due to a bias in the satellite spot ratio between nights, as it is unlikely all three planets brightened in the same way. In Figure \ref{fig:hr8799_spec_fullnight}, the night 1 spectrum of HR 8799 e has enhanced flux at 2.3~$\mu$m that is not seen in night 2 and the other planets. We believe this is likely spurious, and possibly caused by imperfect speckle subtraction at the location of the planet. We will next investigate the stability of the satellite spots to use for spectro-photometric calibration and whether there are systematics to account for.

\section{Satellite Spot Analysis} \label{sec:satspots}

\begin{figure}
    \centering
    \includegraphics[width=0.45\textwidth]{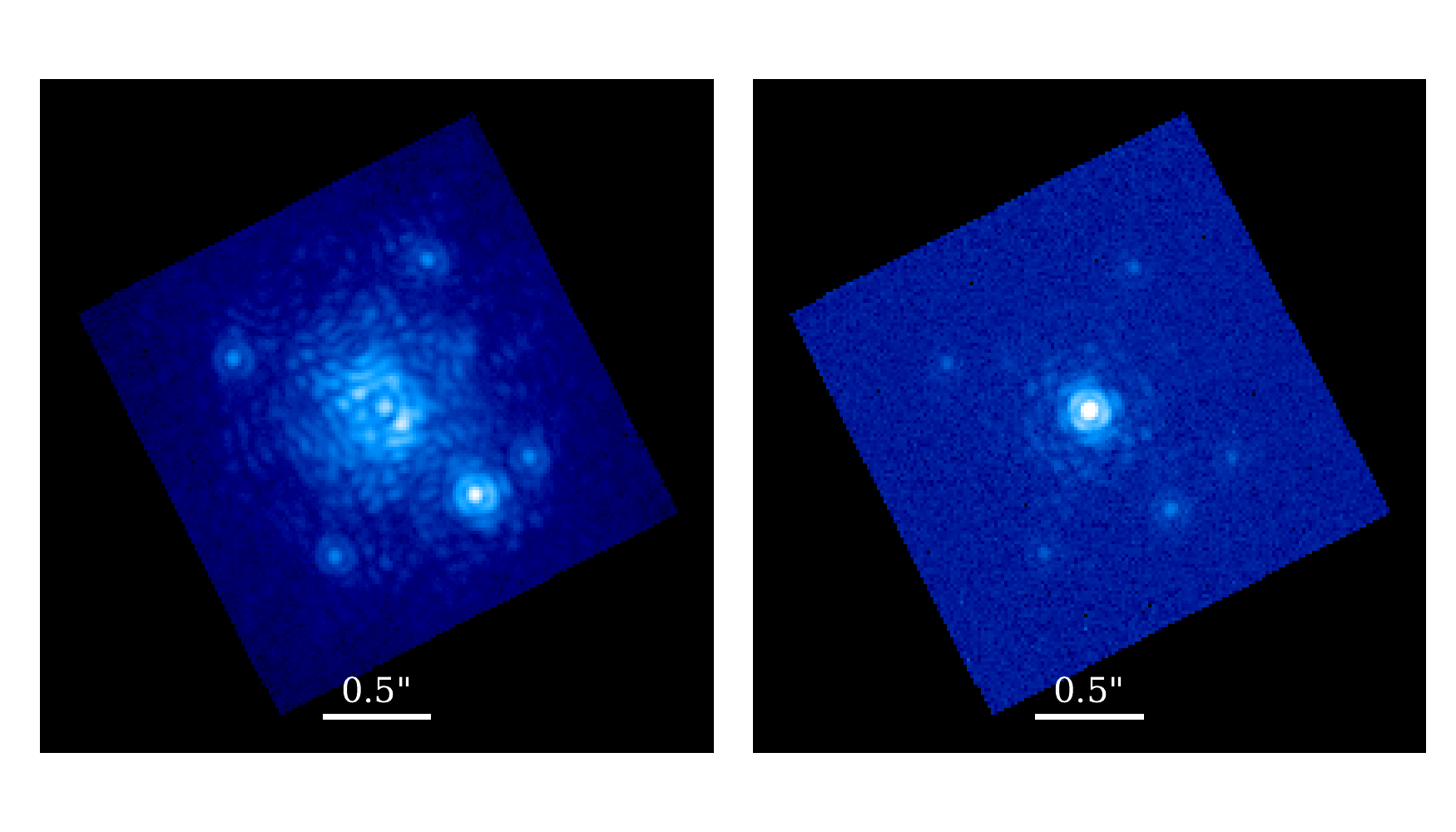}
    \caption{Images of the calibration binary HD 187003. On the left, the primary star is occulted behind the coronagraph, leaving the secondary star and the four satellite spots created by the primary star visible in the image. On the right, the coronagraph is removed and the stars are placed behind a neutral density filter to avoid saturation. Note that four satellite spots, with amplitudes four times as large as the image on the left, are also visible in this image, but not used in the analysis. Both images correspond to the 1.74 $\mu$m channel. }
    \label{fig:satspot_imgs}
\end{figure}

Due to variable adaptive optics correction and unknown atmospheric transmission, simultaneously imaged reference stars are needed to photometrically calibrate data on the HR 8799 planets. With a 2" field of view, the only reference star in the CHARIS data is HR 8799 itself, which is blocked by the coronagraph. Satellite spots are essential for spectro-photometric calibration of this kind of coronagraphic data \citep{Marois2006b,Sivaramakrishnan2006}. Unlike other instruments that use a static pupil mask or deformable mirror offset, the spots used here were generated by placing a fast modulating sine pattern on the deformable mirror to make them incoherent with the speckle field \citep{Jovanovic2015}. We characterize the stability of the SCExAO satellite spots using the data on the HD 187003{, which harbors a companion at $\sim$0.6~arcsec}. We focus on this data because it allows us to calibrate out most effects due to changing atmospheric transmission or adaptive optics correction, since they should affect both components in the same way. 

To our knowledge, HD 187003 is not a known variable star. Even though the primary is a spectroscopic binary consisting of two equal-mass G-type stars, it is not an eclipsing system \citep{Griffin2001}. Typical G-type stars have variability observed at the mmag-level \citep{Ciardi2011}, which is below the 1-10\% variability amplitudes that we can concerned with in this work. {The companion at 0.6~arcsec is an early M dwarf based on our measured $K$-band flux ratio of $\sim$1:30 relative to the unresolved twin G-type stars. Based on \citet{Ciardi2011}, we expect the variability of M-dwarfs to be below 10~mmag, which again is below the 1\% variability amplitudes we are concerned with. For the rest of this paper, we will call this 0.6~arcsec companion the ``binary companion" or ``secondary" for simplicity, since the fact the primary is an unresolved spectroscopic binary can be safely ignored. }

We performed a similar data analysis routine as what was done for measuring the photometry of the planets in each image (see Section \ref{sec:data}). We performed dark subtraction of the raw 2D images, extracted the microlens spectra to form a 3D spectral data cube using the CHARIS data reduction pipeline, and then flat fielded the 3D spectral data cube using data from a flat field lamp taken after the night. This results in one spectral cube per exposure. In Figure \ref{fig:satspot_imgs}, we show representative images of the binary used in the calibration analysis after we have reconstructed them into spectral data cubes. We note that for the exposures taken with the neutral density filter and without the coronagraph, we could not flat field that data since the the flat field exposures were taken with the focal plane mask in that blocks the flat field lamp at the center of the focal plane where the primary star happened to lie in the images. This affects the computed flux ratio of the binary companion at the percent-level, which only affects the photometric calibration at the percent-level. We find later in this section that the photometric calibration of the satellite spots is uncertain beyond this level due to other factors, so the effect is negligible.

\subsection{Binary for Satellite Spot Characterization}\label{sec:satspots-eqn}

\begin{figure}
    \centering
    \includegraphics[width=0.48\textwidth]{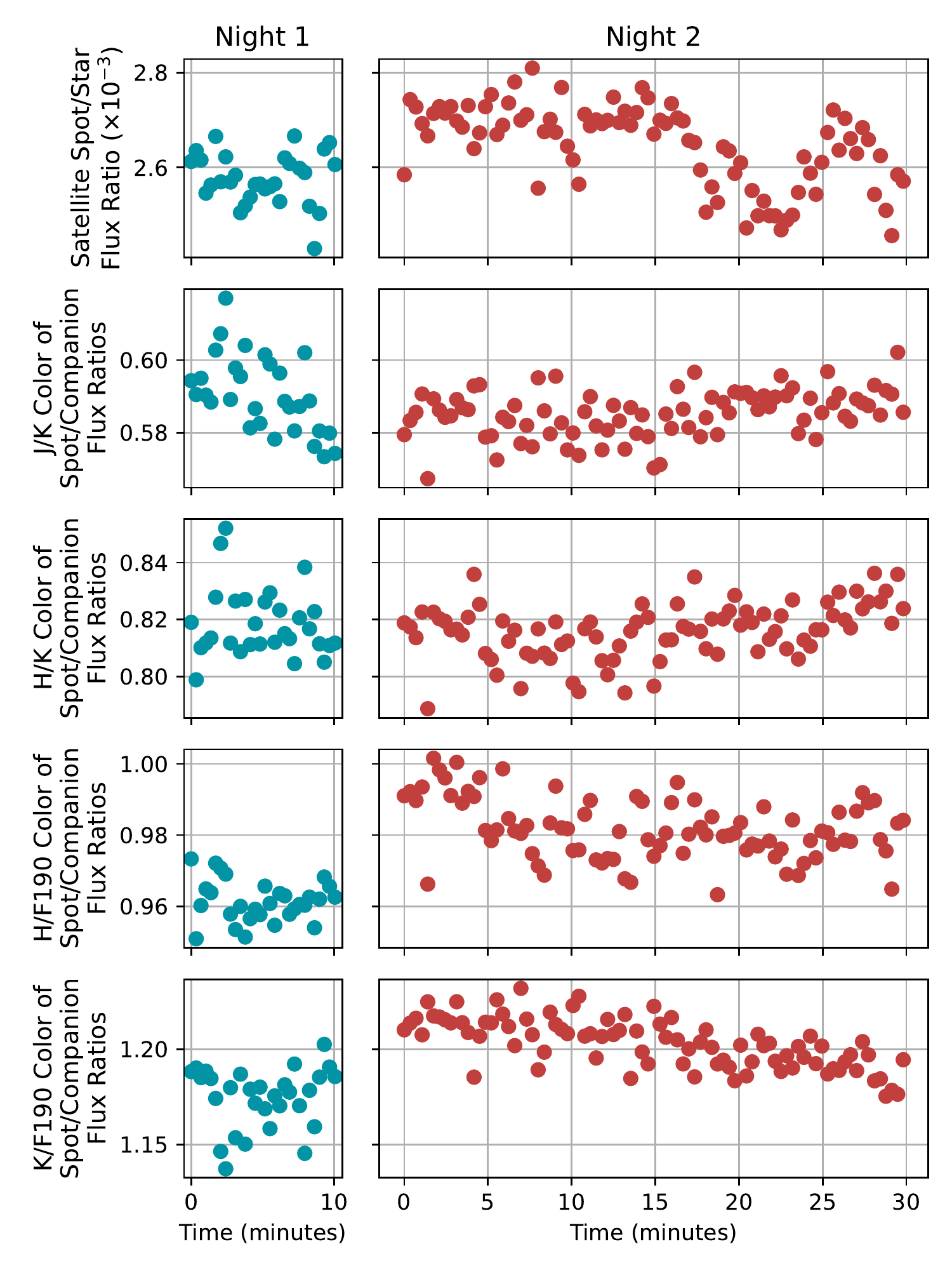}
    \caption{Plots of the photometric and spectroscopic stability of satellite spots as calibrators. The top row shows the photometric stability of the satellite spots (i.e., how fixed is their flux ratio relative to the primary star as described in Equation \ref{eqn:fr-basic}) and the bottom four rows show spectral stability of different colors of the satellite spot relative to the companion (left hand side of Equation \ref{eqn:fr-phot}). Left column shows the data from the first night, and the right column shows the data from the second night. See Section \ref{sec:satspots-eqn} for filter definitions including the F190 filter centered at the 1.9~$\mu$m water absorption band. }
    \label{fig:satspot_time}
\end{figure}

To assess the stability of the satellite spots both photometrically and spectrally over time, we computed the flux ratio between the binary companion and the four satellite spots in each individual frame (each wavelength channel of each spectral datacube). We used the following relations to define our stability tests. Let us define the flux density of the primary and secondary star as a function of wavelength as $F_{A}(\lambda)$ and $F_{B}(\lambda)$ respectively. Then the flux ratio of the secondary to the primary is 
\begin{equation}
    FR_{B,A}(\lambda) = F_{B}(\lambda)/F_{A}(\lambda).
\end{equation}
The satellite spots (labeled with a subscript $s$) are at a given flux ratio relative to the primary star at each wavelength ($FR_{s,A}(\lambda)$) so that their flux density is
\begin{equation}
    F_{s}(\lambda) = FR_{s,A}(\lambda) F_{A}(\lambda).
\end{equation}
Thus, the flux ratio of the secondary relative to the primary can be written as
\begin{equation}\label{eqn:fr-basic}
    FR_{B,A} = \frac{F_{B}}{F_{s}/FR_{s,A}} =  FR_{s,A}/FR_{s, B} 
\end{equation}
where each term depends on wavelength but we have removed the explicit wavelength dependence from the equations. Here $FR_{s, B}$ is the flux ratio of the satellite spots relative to the secondary, which we can directly measure. We also make the assumption that the flux ratio of the secondary relative to the primary is constant, since neither star is known to be significantly variable. Thus, our assumption is that if we measure variations in the flux ratio of the secondary relative to the satellite spots, this is linearly proportional to instabilities in $FR_{s,A}$, the amplitude of the satellite spots in our images.

From just the binary to satellite spot flux ratio as a function of wavelength, we can assess the spectral stability of the satellite spots. Since we assume $FR_{B,A}$ does not change in time, then $FR_{B,A}(\lambda_1)/FR_{B,A}(\lambda_2) = const$ for any two wavelengths $\lambda_1$ and $\lambda_2$. Expanding this ratio of flux ratios and moving terms around so that $\lambda_1$ is only in the numerator, we get that
\begin{equation}
    \frac{FR_{s,A}(\lambda_1)}{FR_{s,A}(\lambda_2)} = \frac{FR_{B,A}(\lambda_1)}{FR_{B,A}(\lambda_2)}\frac{FR_{s,B}(\lambda_1) }{FR_{s, B}(\lambda_2) }.
\end{equation}
This means that any changes in $FR_{s,B}(\lambda_2)/FR_{s,B}(\lambda_1)$ is linearly proportional to changes in $FR_{s,A}(\lambda_2)/FR_{s,A}(\lambda_1)$. Changes in the latter should be due to changes in the spectral stability of the satellite spots, so $FR_{s,B}(\lambda_2)/FR_{s,B}(\lambda_1)$ allows us to assess the spectral stability of the satellite spots.

We focused on testing the spectral stability in the ratio of photometry in different bands (i.e., colors) so we will take the mean of the flux ratio over each band. Thus, what we are comparing is
\begin{equation}\label{eqn:fr-phot}
    \frac{\sum_i FR_{s,B}(\lambda_i) }{\sum_j FR_{s,B}(\lambda_j)} = \frac{ \sum_i FR_{s,A}(\lambda_i)/FR_{B,A}(\lambda_i)}{ \sum_j FR_{s,A}(\lambda_j) / FR_{B,A}(\lambda_j) }
\end{equation}
where for each of the bands, we sum over all wavelength indices (either $i$ or $j$) that are defined to be in that band. We also assumed each band has the same number of wavelength channels so that the normalization by number of wavelengths for the mean cancels out. We defined bands from the CHARIS spectra that are roughly the $J$, $H$, and $K$ bands, as well as a custom ``F190" band that is centered on the water absorption band at roughly 1.9~$\mu$m. For this work, we defined the bands as follows: $J$ band is wavelength channels 1 through 4 inclusive (1.20-1.33 $\mu$m); $H$ band is channels 9 through 12 inclusive (1.58-1.74 $\mu$m); the custom F190 band is channels 13 through 16 inclusive (1.80-2.00 $\mu$m); $K$ band is channels 17 through 20 inclusive (2.07-2.29 $\mu$m). All indices are assumed to start from channel 0. We note that these are not the exact definitions for these photometric bands, but are a convenient approximation for this work.

\subsection{Satellite Spot Photometry}\label{sec:satspot-phot}

We investigated the photometric stability of the satellite spots and measured their flux ratio, $FR_{s,A}(\lambda)$, for photometric calibration. To do this, we need to measure $FR_{B,A}$ amd $FR_{s,B}$ as shown in Equation \ref{eqn:fr-basic}.
As we used PSF fitting for the HR 8799 planets, we used PSF fitting here too to measure these relative flux ratios. First for $FR_{s,B}$, since the binary companion is brighter than the satellite spots, we used it as the PSF model that we fit to the satellite spots. To extract the binary companion PSF, we performed centroiding on the companion in each frame by fitting it to a 2-D Airy PSF and then extracting out a 21x21 pixel box based on the centroid. This resulted in a 21x21 PSF of the binary companion for each frame to fit to each of the four satellite spots. We performed the same Airy PSF centroiding to find the approximate location of each satellite spot, and defined a 15x15 pixel fitting region around each satellite spot. We performed a least-squares optimization using the Levenberg-Marquardt algorithm to find the best x and y shifts and the best flux ratio scaling of the companion PSF to match each satellite spot. We took the mean of the four flux ratios obtained from this fit to be the flux ratio between the binary companion and the satellite spots at that given wavelength and time. We averaged over the respective wavelength channels that we have defined in each band to obtain average flux ratios in our $J$, $H$, F190, and $K$ bands for each exposure. 

We then computed the flux ratio between the primary and secondary star, $FR_{B,A}(\lambda)$. In the images where the primary is behind the neutral density filter rather than the occulting masking, we perform the same PSF fitting routine to measure the binary flux ratio as we had done to measure the flux ratio between the secondary and the satellite spots in the previous spectral stability analysis. With $FR_{B,A}(\lambda)$ measured from the unocculted images, we can use the measured $FR_{s, B}$ in each frame of coronagraphic data to derive $FR_{s,A}$ following the relationship in Equation \ref{eqn:fr-basic}.

We averaged the flux ratio of the four satellite spots in each wavelength channel of each exposure. Because the satellite spot flux ratios intrinsically scale $\propto \lambda^{-2}$, we rescaled all of them to a fiducial wavelength $\lambda_0$ by multiplying each flux ratio by $\lambda^2/\lambda_0^2$. We picked $\lambda_0 = 1.55 \mu$m to be consistent with previous work \citep{Currie2018}. We then averaged the rescaled satellite spots across all wavelengths in a single exposure, and plotted the measured satellite spot flux ratio at 1.55 $\mu$m for each exposure in the bottom row of Figure \ref{fig:satspot_time} to assess their photometric stability.

We see drifts up to 10\% in the satellite spot flux ratio over the course of half an hour. One possible cause could be due to the fact that SCExAO sees on the wavefront sensor (WFS) the sine pattern from the deformable mirror (DM) that creates the satellite spots. Even though that the SCExAO control loop is designed to filter out this signal, the filtering could be imperfect and could depend on what the signal of atmospheric turbulence looks like on the WFS. Another control-related cause could be imperfect DM calibration, where the voltage to physical displacement conversion is incorrect and the amplitude of the sine wave that creates the satellite spots changes as the shape of the DM changes to correct for atmospheric turbulence. Alternatively, the quasi-static speckles at the location as the satellite spots may be evolving over time as the adaptive optics correction changes. Since speckles are also $\lambda$/D in size, it is difficult to disentangle these two signals when measuring the satellite spot fluxes. We did not alternate turning the spots on and off as has been done in other SCExAO observations to remove quasi-static speckles at the location of the satellite spots \citep{Sahoo2020}. All three processes imply that the flux ratio of the satellite spots could change as conditions and turbulence change. We measured a 2.1\% and 3.2\% scatter in the satellite spot flux ratios as a function of time in night 1 and night 2 respectively.

We found an average satellite spot flux ratio at 1.55 $\mu$m of $2.58 \times 10^{-3}$ and $2.65 \times 10^{-3}$ for night 1 and night 2 respectively, a 3\% difference that is much smaller than the drift within night 2. Despite this, the fact the flux ratio is very correlated in time makes it suspect that the flux ratio could stay constant over the course of a whole night. It may be that we observed the binary at similar hour angles on consecutive nights where the atmospheric conditions were both consistent and good. Data taken at other hour angles or in changing conditions may produce a different satellite spot ratio. As seen in Figure \ref{fig:satspot_time}, the 30 minute sequence on night 2 has a possible downwards trend in the satellite spot flux ratio over the full time baseline, indicating there could be longer timescale variations that are not measured in this data. The correlated fluctuations may not average out when taken in more heterogeneous conditions. This may explain why our derived satellite spot flux ratio is also different by 3\% from the $2.72 \times 10^{-3}$ reported by \citet{Currie2018} and why our measured HR 8799 planet spectra are all brighter on the second night (Figure \ref{fig:hr8799_spec_fullnight}). These data only characterize the stability on 30-minute and 24-hour timescales. Future observations that observe binary stars over more varied timescales are necessary to better characterize this instability.

\subsection{Satellite Spot Spectral Stability}\label{sec:satspot-spec}

We also investigate the spectral stability of the satellite spots by looking at the stability of the color of the flux ratio between the satellite spots and the binary companion.
In Figure \ref{fig:satspot_time}. we plot the colors of the satellite spot to secondary flux ratios (i.e., quantity in the left hand side of Equation \ref{eqn:fr-phot}) in the following combinations of bands: $J$/$K$, $H$/$K$, $H$/F190, $K$/F190. 

In these time-series shown in Figure \ref{fig:satspot_time}, we see some 2\% drifts on the 10-minute timescales as well as on the night-to-night timescales, but they all are much smaller than the photometric drifts. All four ratios show drifts of 1-2\% on the 10- to 30-minute timescales. The $J$/$K$ and $H$/$K$ ratios show negligible ($< 1$\%) variations between the two nights: $J$/$K$ ratio is $0.591 \pm 0.010$ on night 1 and $0.585\pm 0.006$ on night 2 and $H$/$K$ ratio is $0.819 \pm 0.012$ on night 1 and $0.816 \pm 0.010$ on night 2 (the error bars here represent the scatter between exposures within a night). However, the two ratios involving the F190 filter, which is strongly affected by telluric absorption, do see a significant $\sim$2\% shift between the two nights: the $H$/F190 ratio is $0.961 \pm 0.006$ on night 1 and $0.982 \pm 0.009$ on night 2; the $K$/F190 ratio is $1.175 \pm 0.016$ on night 1 and $1.203 \pm 0.013$ on night 2. Given the 2\% variations in the satellites spot colors exist on the minutes to day timescales, we find a systematic floor in measuring the colors of any exoplanet of 2\%. Anisoplanatism could be one possible cause of the drift: the satellite spots probe the atmospheric turbulence of the primary star, which is also used for adaptive optics correction, but the light of the binary (and any planet companions) is separated by 400 to 1000~mas and travels through a slightly different patch of atmosphere and suffers slightly degraded adaptive optics correction. The amount of degradation is time variable depending on the structure of atmospheric turbulence and is also wavelength-dependent as shorter wavelengths are generally more affected. We note that we did not probe drifts on the hour timescales from this data, so there could be additional drifts at those timescales, especially if anisoplanatism changing with changing airmass is a dominant effect. Still, these systematic drifts are only observed at the percent-level.

\section{Atmospheric Variability}\label{sec:var}
\subsection{Photometric Variability}\label{sec:phot-var}

\begin{figure*}
    \centering
    \includegraphics[width=0.95\textwidth]{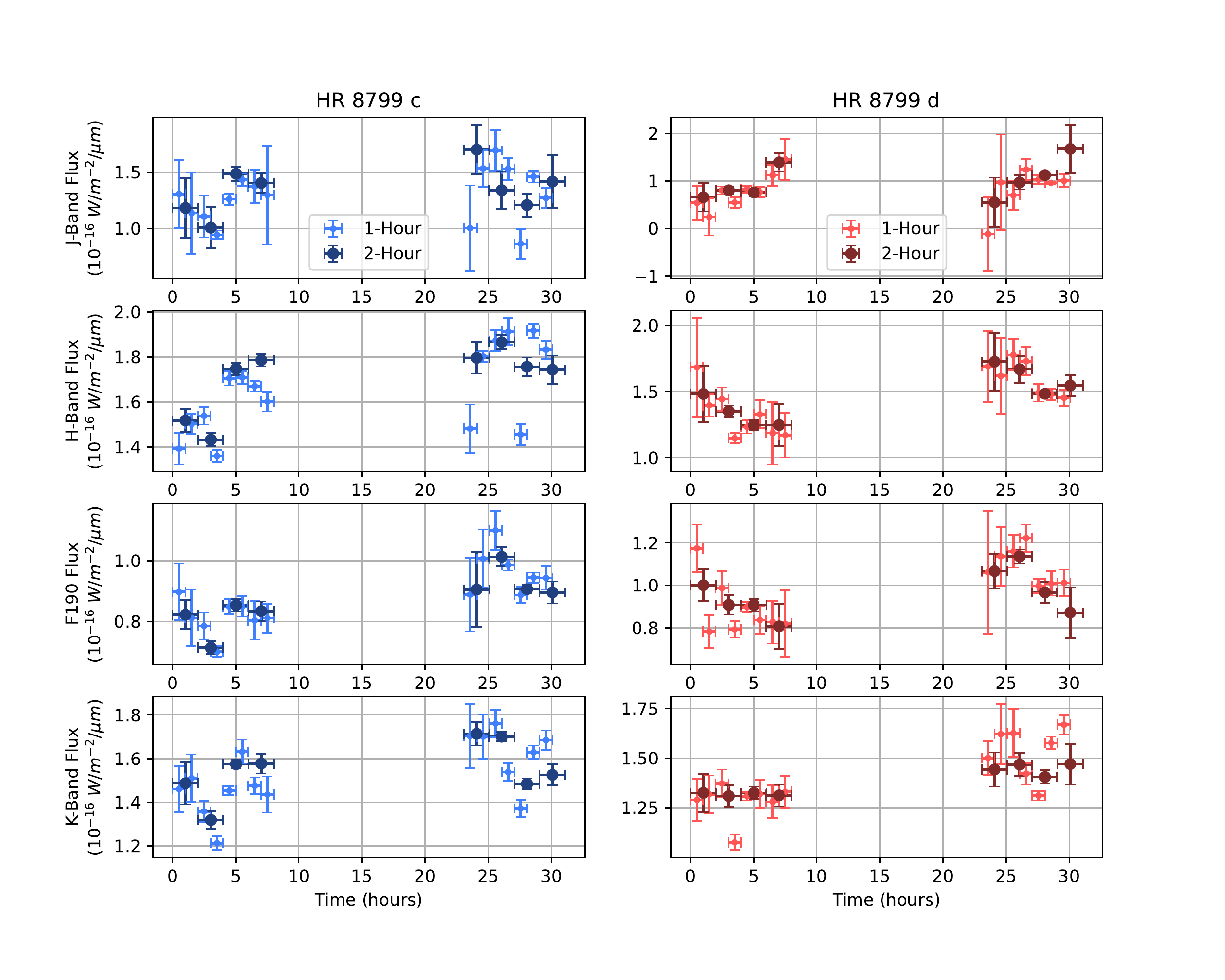}
    \caption{Time-series of the flux of HR 8799 c (left column) and HR 8799 d (right column) integrated over $J$, $H$, F190, and $K$ band (see Section \ref{sec:satspots-eqn} for the definitions of these bands). In each panel, photometry from both 1- and 2-hour bins are plotted. Error bars represent statistical errors and do  not incorporate biases due to photometric calibration. The 1-hour and 2-hour photometric time-series in this figure are available as the Data behind the Figure.  }
    \label{fig:cd_phot_time}
\end{figure*}

\begin{figure*}
    \centering
    \includegraphics[width=0.95\textwidth]{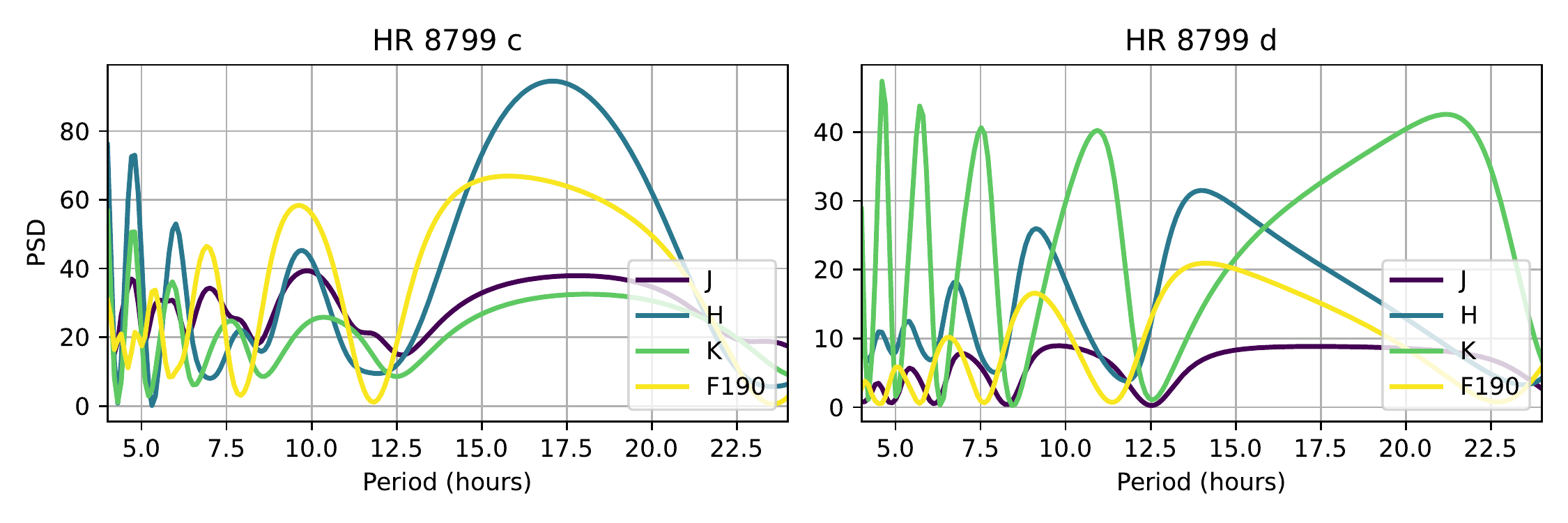}
    \includegraphics[width=0.95\textwidth]{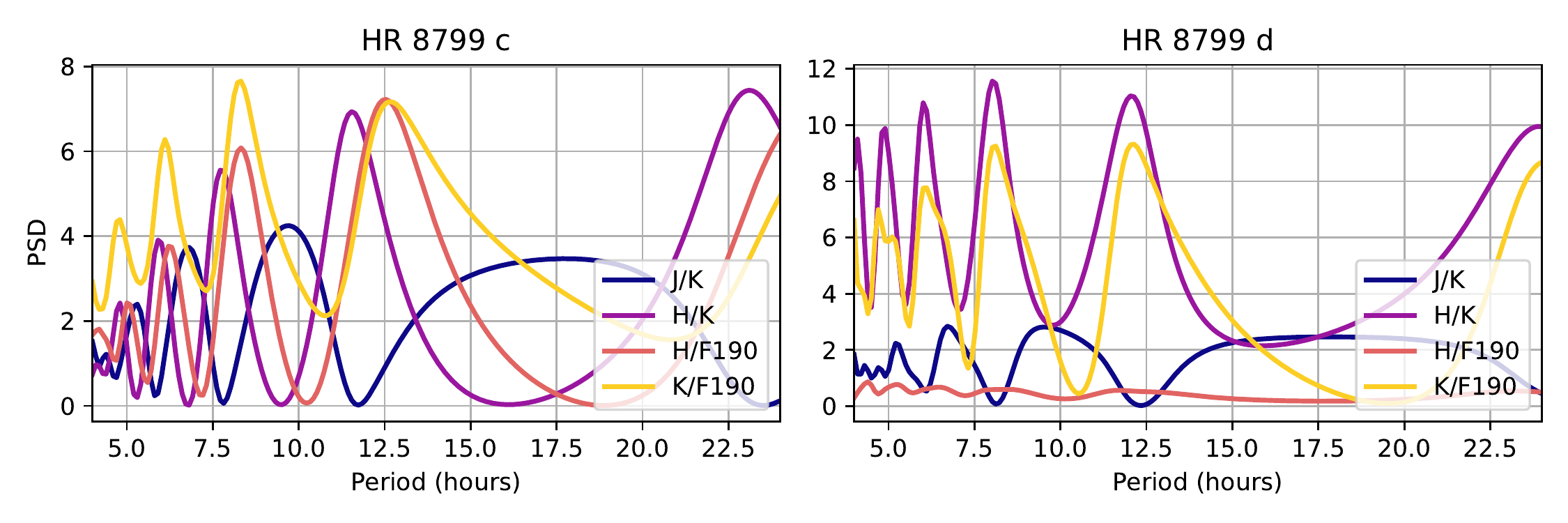}
    \caption{Lomb-Scargle periodograms for HR 8799 c (left) and HR 8799 d (right) using both the photometric time-series (top) and flux ratio time-series (bottom). No significant peaks are found in any of the configurations. }
    \label{fig:periodograms}
\end{figure*}

\begin{deluxetable*}{c|c|c|c|c|c}
\tablecaption{Peak photometric variability periods from periodogram analysis  \label{table:phot_peaks}}
\tablehead{
Planet & Band & Period (hours) & Amplitude (W/m$^2$/$\mu$m) & Amplitude (\%) &  False Alarm Probability (\%)
}
\startdata
c & J & 4.0 & $2.5 \times 10^{-17}$ & 21 & 50 \\
c & H & 17.1 & $1.8 \times 10^{-17}$ & 11 & 45 \\
c & K & 4.0 & $1.9 \times 10^{-17}$ & 13 & 32 \\
c & F190 & 15.8 & $1.1 \times 10^{-17}$ & 13 & 37 \\
\hline
d & J & 9.8 & $1.5 \times 10^{-17}$ & 18 & 34 \\
d & H & 13.9 & $2.4 \times 10^{-17}$ & 17 & 22 \\
d & K & 4.6 & $1.8 \times 10^{-17}$ & 13 & 52 \\
d & F190 & 14.1 & $1.3 \times 10^{-17}$ & 14 & 32 
\enddata
\end{deluxetable*}

\begin{figure}
    \centering
    \includegraphics[width=0.46\textwidth]{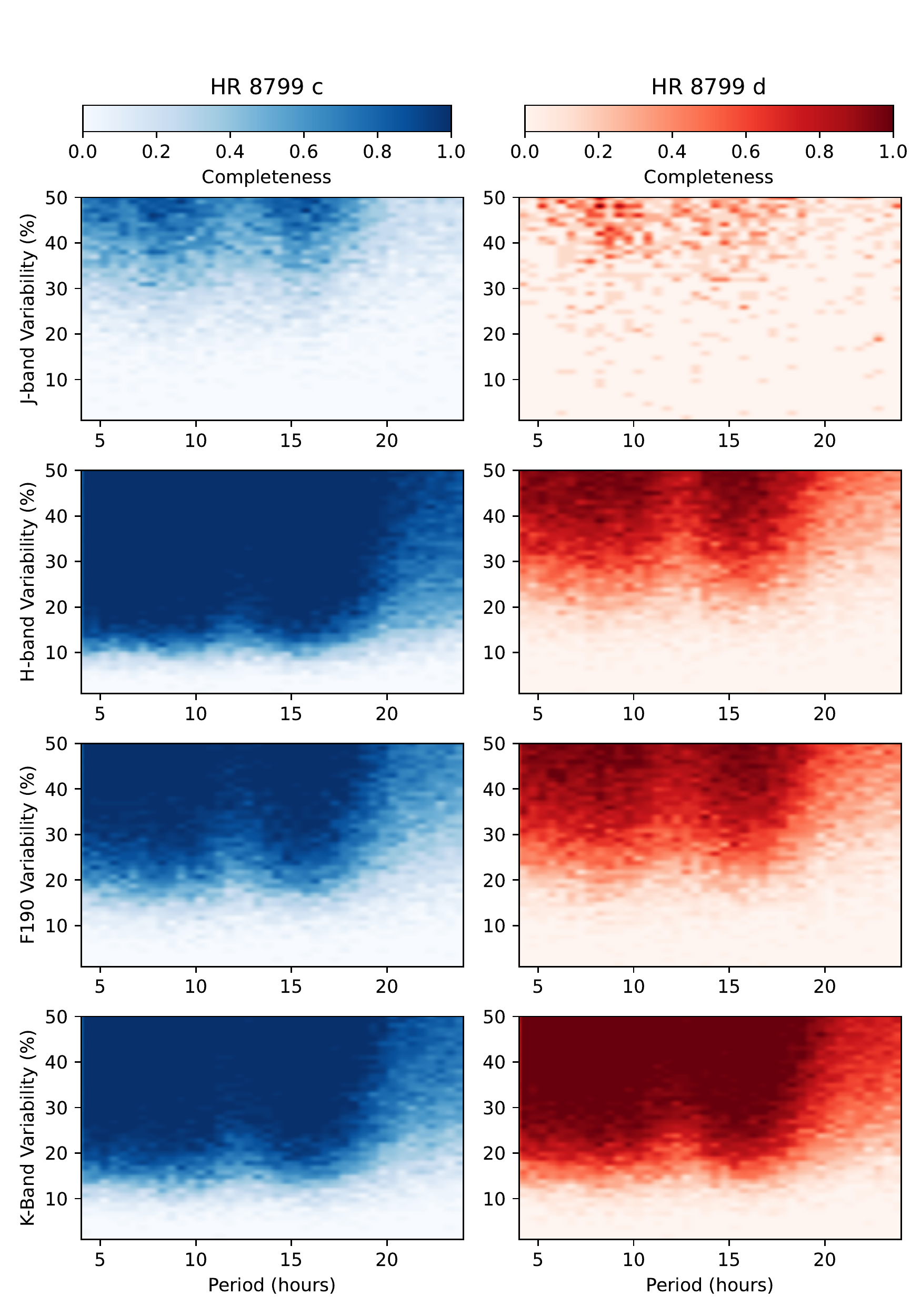}
    \caption{Completeness maps to photometric variability in four bands at a given fractional amplitude and period for HR 8799 c (left) and HR 8799 d (right). }
    \label{fig:complete-phot}
\end{figure}

Since most brown dwarf variability studies have been looking at photometric bands \citep[e.g.,][]{Radigan2014, Artigau2018}, we will spectrally bin the flux of the planets into various photometric bands and look at the corresponding time series plots to avoid the complexity of the flux changing in 22 spectral channels. We define our own $J$, $H$,  $K$ and F190 bands as we detailed in Section \ref{sec:satspots-eqn}. Each band consists of four spectral channels where the flux of the planet is brightest. The uncertainties for each datapoint were estimated based on the scatter in the simulated planet fluxes that were binned into the same spectral bands. We plot the time series for HR 8799 c and d in Figure \ref{fig:cd_phot_time}. The SNR on HR 8799 e was too poor. In Figure \ref{fig:cd_phot_time}, we see that the fluxes on the second night are higher on average compared to the first night, consistent with the global scaling seen in Figure \ref{fig:hr8799_spec_fullnight} which we hypothesize could due to a bias in the satellite spot flux ratios between the two nights. Ignoring this global offset between nights, we see that the photometry of each planet is correlated between bands, but do not appear to be strongly correlated to the other planet. 

This implies that the variability within a night we see is not due to the instability of the satellite spots, as it should affect both planets in the same way. Instead, this photometric behavior could be due to real changes in the planet's atmosphere or unaccounted-for errors in the spectral extraction. The amplitude of the flux variability is quite large (20\%) and near the limit of what is seen for comparable low mass objects \citep{Bowler2020}. Systematic errors in the spectral extraction process are somewhat accounted for with the simulated planet injection analysis described in Section \ref{sec:spectral_extraction}, which essentially injected planets with a flat light curve and made corrections so that the output spectrophotometry of the simulated planets was flat, on average. However, this process does not account for all possible systematics. One remaining systematic is a mismatch between the true planet PSF and the PSF we derived from the satellite spots, as the satellite spots are radially elongated due to the finite spectral bandwidth of each channel in the CHARIS low-resolution mode. A more detailed study of the accuracy of spectrophotometry in coronagraphic images, which does not exist in the literature to the best of our knowledge, may help determine how relevant this concern is.

To formally assess the light curves for any significant periodic variability, we constructed Lomb-Scargle periodograms for the time-series of each planet in each photometric band using the implementation in \texttt{astropy} \citep{Astropy2013,Astropy2018}. We found the results from the one-hour chunks to be more sensitive from a bootstrap analysis of the false alarm probability, so we will focus only on those results. The Lomb-Scargle periodograms are plotted in Figure \ref{fig:periodograms}, and it is clear there is no significant peak in any band. The most significant peak for each band is listed in Table \ref{table:phot_peaks} along with its false alarm probability. All the false alarm probabilities are greater than 5\%, so none of them should be considered significant. We note that this analysis assumes the variability is sinusoidal, and ignores variability in the variability, although this is likely a second-order effect: a primary peak corresponding to the rotation period of the object should still be discernible. 

To quantify the sensitivity of our data to periodic variability signals, we injected sinusoidal signals into the one-hour-chunk data at a variety of different amplitudes and periods. At each amplitude and period, we performed 20 injection and recovery tests, varying the phase of the sinusoid in each injection. In each test, the signal was injected with noise drawn from Gaussian distributions with widths equal to the quadratic sum of the statistical errors computed from the data and a 3\% systematic error due to time-variable satellite spot calibrations discussed in Section \ref{sec:satspot-phot}. We added the 3\% satellites spot calibration systematic as a random error term as we do not have data to quantify its trend on hour-long timescales. The only measurement with time baselines longer than an hour that we have is the fact that the satellite spot flux ratio changed by 3\% between the two nights. Thus, we caveat the following analysis with the fact our expected sensitivity could be worse due to additional unknown systematics.

After the simulated signal was injected, the time-series was run through the same Lomb-Scargle periodogram analysis above and determined to be a detection if the periodogram found the correct period with a false alarm probability $< 5$\%. We computed the fraction of signals at each variability amplitude and period that were detected to determine how complete we are for that amplitude and period (i.e., the chance we would have detected such a signal in the data). These completeness values were evaluated on a grid of variability amplitudes and periods to form completeness maps. Completeness maps for each photometric band and for both planets are shown in Figure \ref{fig:complete-phot}. In the best-case {for rotation periods between 5 and 18 hours}, we maintain $ > 50$\% completeness for $10\%$ variability amplitudes for HR 8799 c in $H$ band. These results are 2-3x better than \citet{Biller2021}, who also did $H$-band photometric monitoring of HR 8799 c with SPHERE. For HR 8799 d, we only reach down to 30\% variability amplitudes in $H$-band, but these are the first significant constraints on photometric variability for this planet. Our $K$-band sensitivities are $> 50$\% complete to variability amplitudes of 20\% for both planets, and are the first in $K$ band for these planets. However, the patchy cloud models from \citet{Skemer2014} do not predict $> 10\%$ variability in $K$-band, so the lack of variability may be expected. We also note that all of the periodogram peaks listed in Table \ref{table:phot_peaks} have amplitudes and periods that correspond to $\leq$~50\% completeness in our completeness maps. This reinforces the non-detection of photometric variability in this HR 8799 dataset.

\subsection{Color Variability}\label{sec:color-var}

\begin{figure*}
    \centering
    \includegraphics[width=0.95\textwidth]{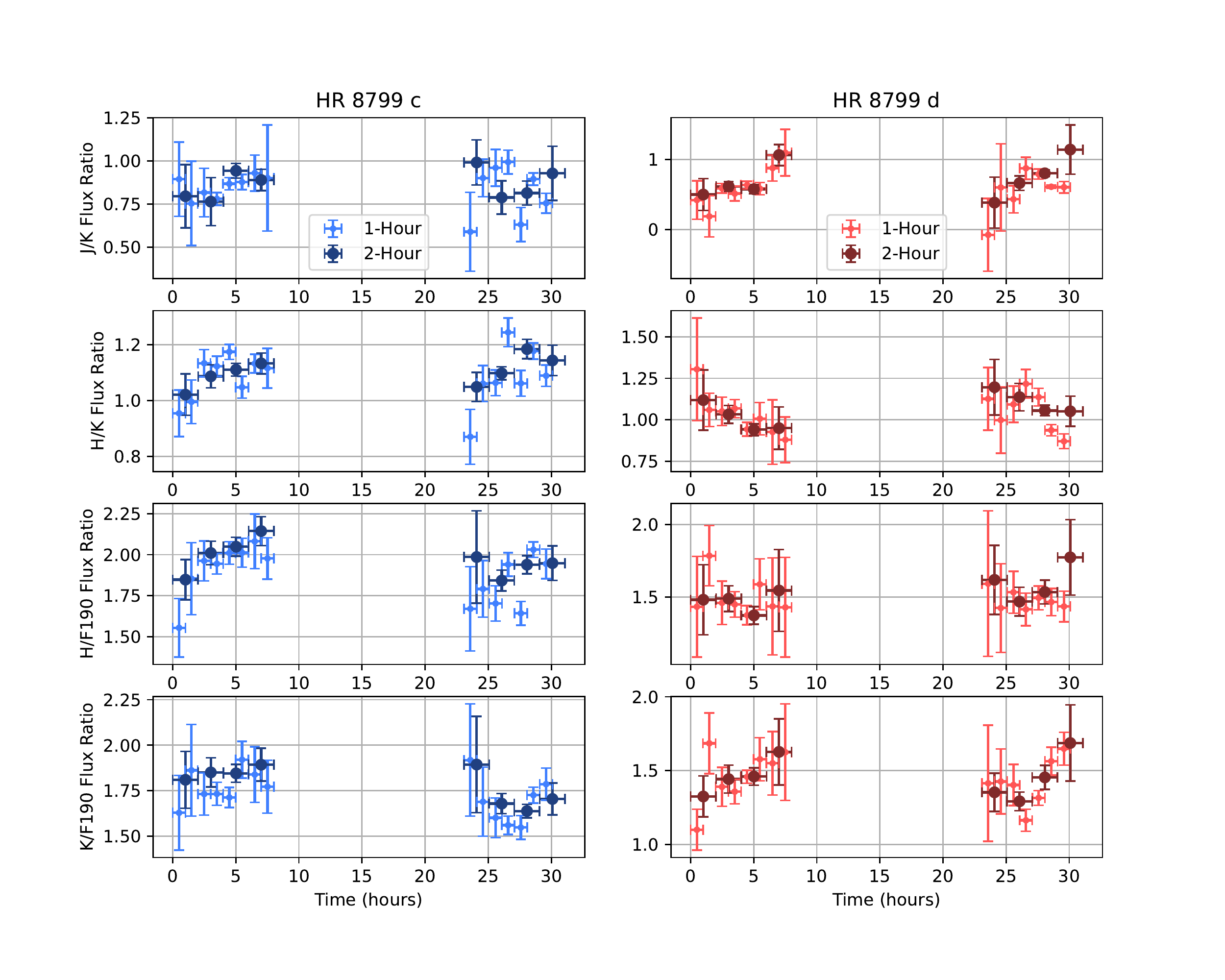}
    \caption{Time-series of the flux ratios between various bands for HR 8799 c (left column) and HR 8799 d (right column). In each panel, photometry from both 1- and 2-hour bins are plotted. Error bars represent statistical errors. The 1-hour and 2-hour flux ratio time-series in this figure are available as the Data behind the Figure. }
    \label{fig:cd_color_time}
\end{figure*}

\begin{deluxetable*}{c|c|c|c|c|c}
\tablecaption{Peak color variability periods from periodogram analysis  \label{table:ratio_peaks}}
\tablehead{
Planet & Bands & Period (hours) & Flux Ratio Amplitude & Amplitude (\%) &  False Alarm Probability (\%)
}
\startdata
c & J/K & 9.7 & 0.09 & 10.8 & 74 \\
c & H/K & 23.1 & 0.34 & 41.5 & 53 \\
c & H/F190 & 12.5 & 0.16 & 8.3 & 60 \\
c & K/F190 & 8.3 & 0.13 & 7.8 & 32 \\
\hline
d & J/K & 6.7 & 0.10 & 15.7 & 61 \\
d & H/K & 8.0 & 0.14 & 13.9 & 30 \\
d & H/F190 & 4.4 & 0.06 & 4.1 & 71 \\
d & K/F190 & 12.1 & 0.24 & 16.0 & 40
\enddata
\end{deluxetable*}

\begin{figure}
    \centering
    \includegraphics[width=0.46\textwidth]{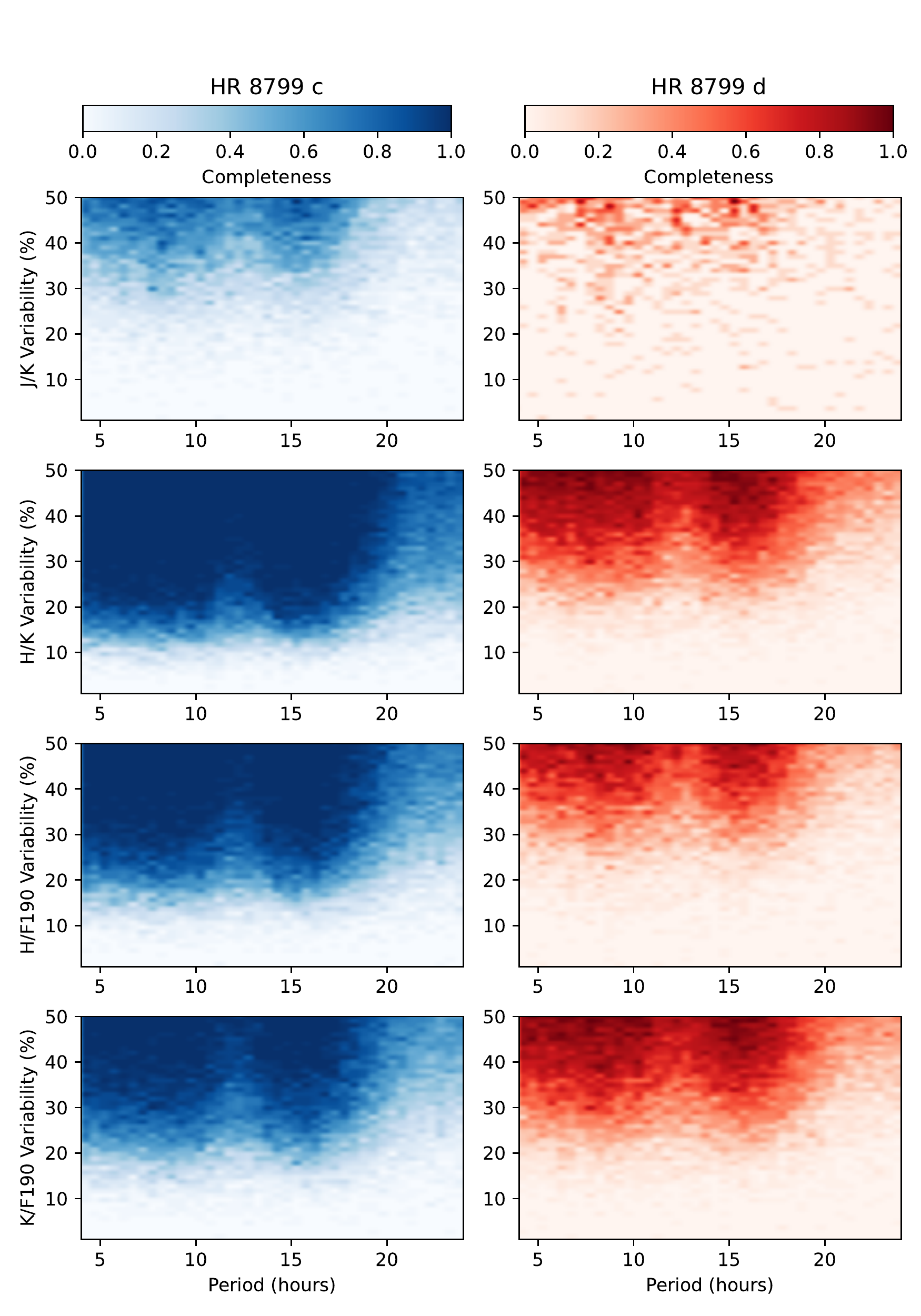}
    \caption{Completeness maps to color variability in the same fashion as Figure \ref{fig:complete-phot}. }
    \label{fig:complete-ratio}
\end{figure}

We also investigated if the broadband colors would be a better diagnostic of variability, as it removes effects such as the global calibration offset between the two nights. There is also scientific motivation for this. The partly cloudy models of the HR 8799 planets presented in \citet{Skemer2014} predict an order of magnitude more variability in $J$ and $H$ band compared to $K$ band. Variability studies of brown dwarfs have also found significantly decreased variability inside the water absorption bands than outside of them \citep{Zhou2020}. Motivated by these previous works, we plot the $J/K$, $H/K$, $H$/F190, and $K$/F190 flux ratios of HR 8799 c and d in Figure \ref{fig:cd_color_time}. We would expect all four to show a similar trend if there is real variability on the planets. 

The colors evolve much more gradually over time than the photometric light curves presented in the previous section, possibly indicating systematic biases are cancelling out. We find that the scatter between adjacent points is comparable to the formal uncertainties denoted by the error bars, unlike in the case of the photometric light curves where outliers were common. We argue from this that the spectral time series data is likely more robust. 

In night 1, we find a that the trends in the HR 8799 c colors correlate with trends in the photometry: all four flux ratios increase as the photometry increases for HR 8799 c. However, no clear trend emerges in night 2, and no consistent trend exists for HR 8799 d. The different changes in color between the two planets argues against these trends due to some spectral instability in the satellite spots as that would affect both planets in the same way. To statistically assess the significance of any periodicity in the signal, we run the same Lomb-Scargle periodogram analysis on the 1-hour time series for the flux ratios. This time, we included a 2\% satellite spot color systematic floor as discussed in Section \ref{sec:satspot-spec}. The periodograms are plotted in Figure \ref{fig:periodograms} and the peak periods are reported in Table \ref{table:ratio_peaks}. Again, no significant peaks are found, so we conclude that we should not read too much into any correlations in the HR 8799 c colors. We use the same injection and recovery test as the previous section to assess our sensitivity to color variability at the range of periods and amplitudes in Figure \ref{fig:complete-ratio}. The most sensitivity flux ratio is the $H$/$K$ ratio, for which we should be sensitive to $\sim$20\% variability amplitudes for HR 8799 c. The completeness maps also confirm that none of the peak periods reported in Table \ref{table:ratio_peaks} are significant as they all reside in areas with $\leq 50$\% completeness.

\subsection{Comparison with other substellar objects}
Most of the near-infrared variability in brown dwarfs and planetary mass companions have been identified with $J$-band photometric monitoring \citep{Radigan2014, Artigau2018}, making it difficult to compare to our constraints due to our poor sensitivity in $J$ band (Figures \ref{fig:complete-phot} and \ref{fig:complete-ratio}). Fortunately, there are a few suitable comparison objects that do have spectral or $H$-band variability, so we will focus on individual comparisons. 

We can compare these limits to the variability of VHS~J1256-1257~b, a wide separation planetary mass companion with similar mass and temperature as the HR 8799 planets \citep{Gauza2015, Bowler2020, Zhou2020}. Thus we expect similar variability behavior between the HR 8799 planets and VHS~J1256-1257~b, modulo viewing angle. The near-infrared spectra from the Hubble Space Telescope indicates large variability with amplitudes of 10\% in $H$ band 13\% in $J$ band \citep{Bowler2020}. Since their spectrum does not cover $K$ band and we have poor constraints bluer of $H$ band, the comparisons of empirical data is not very constraining.  However, their model of variability \citep{Zhou2020} predicts variability amplitudes of $\sim$5\% in the F190 band and $\sim$7\% in the $K$ band (see their Figure 10, noting they are plotting peak-to-valley variability). Given the $\sim$2x smaller variability amplitudes in the F190 and $K$ bands, we expect $\sim$10\% variability amplitudes for the $H$, $H$/F190, and $H$/$K$ time series of the HR 8799 planets if they are viewed edge-on like VHS~J1256-1257~b appears to be \citep{Zhou2020}. While we are not very sensitivity to $H$/F190 or $H$/$K$ spectral variability at these amplitudes, our $H$-band photometric variability sensitivity is $> 50\%$ complete to such an amplitude for periods $< 20$~hours. Thus, it is unlikely that HR 8799 c is viewed equator-on if it is as strongly variable as VHS~J1256-1256~b. 

Given the current constraint on the line-of-sight inclination of the orbital plane of the four planets is $26.8^\circ \pm 2.3^\circ$ \citep{Wang2018}, the planets would need to have a high obliquity if we were viewing them equator-on. If the HR 8799 planets are not viewed equator-on, then the variability amplitudes would be much lower. The free-floating planetary mass companion PSO~J318.5-22 also has a similar temperature and mass as the HR 8799 planets \citep{Allers2016}, and has a spin axis inclination of $61^\circ \pm 17^\circ$. It only experiences a 2\% variability amplitude in the $H$ band \citep{Biller2018} that is well-below our detection limits. However, our non-detection alone does not rule out the possibility of the HR 8799 planets being equator-on, as they may not have the same high-amplitude variability as VHS~J1256-1256~b. Assuming a simple and naive $\sin(i)$ dependence of variability amplitude on inclination, the $H$-band variability of PSO~J318.5-22 would still be $< 3$\% if viewed equator-on, and an equivalent variability would be undetected in our data for the HR 8799 planets.

Hotter temperature objects also are shown to have reduced variability in the $H$ band, but the variability in the water bands are not suppressed, indicating their patchy clouds reside high up in the atmosphere \citep{Yang2015,Lew2016}. Given they are spectrally dissimilar to the HR 8799 planets, it is unlikely that the exact same case is happening for the HR 8799 planets. However, we mention it as a possible reason for spectral variability to be suppressed when comparing $H$ and $K$ band to the F190 band in our analysis. 

\section{Spectral Modeling}\label{sec:atm-fits}

\begin{figure*}
    \centering
    \includegraphics[width=0.8\textwidth]{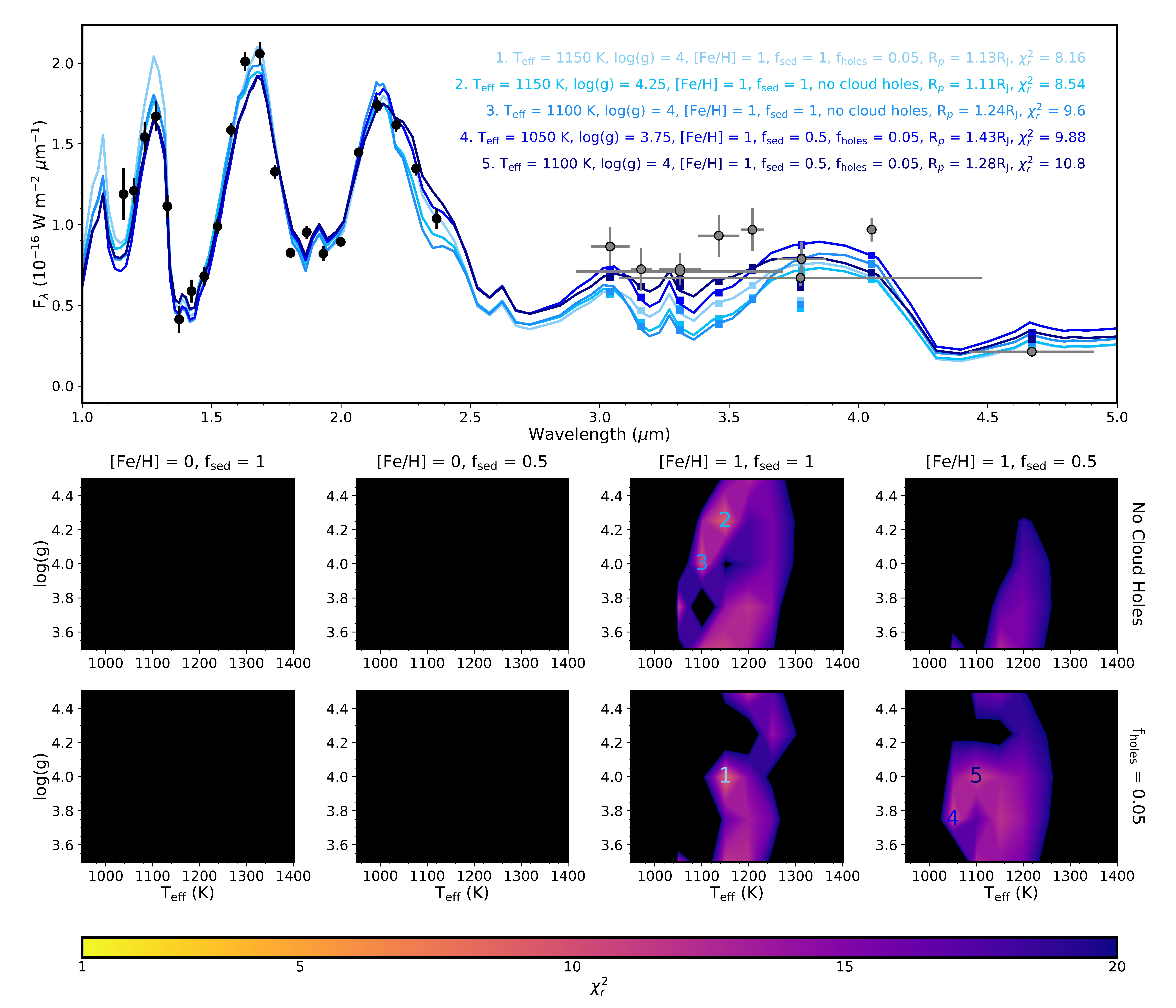}
    \caption{Best fit model emission spectra (top) and reduced $\chi^2$ maps (bottom) for HR 8799~c. The five models with the lowest reduced $\chi^2$ are shown in shades of blue, with their best fit planet radii and reduced $\chi^2$ given, and compared to data from this work (black circles) and previous works \citep[grey circles; see][and references therein]{Bonnefoy2016}. The blue squares show the filter-averaged model values. The locations of the five models in the reduced $\chi^2$ maps are indicated by the corresponding numbers.  }
    \label{fig:hr8799c_model}
\end{figure*}

\begin{figure*}
    \centering
    \includegraphics[width=0.8\textwidth]{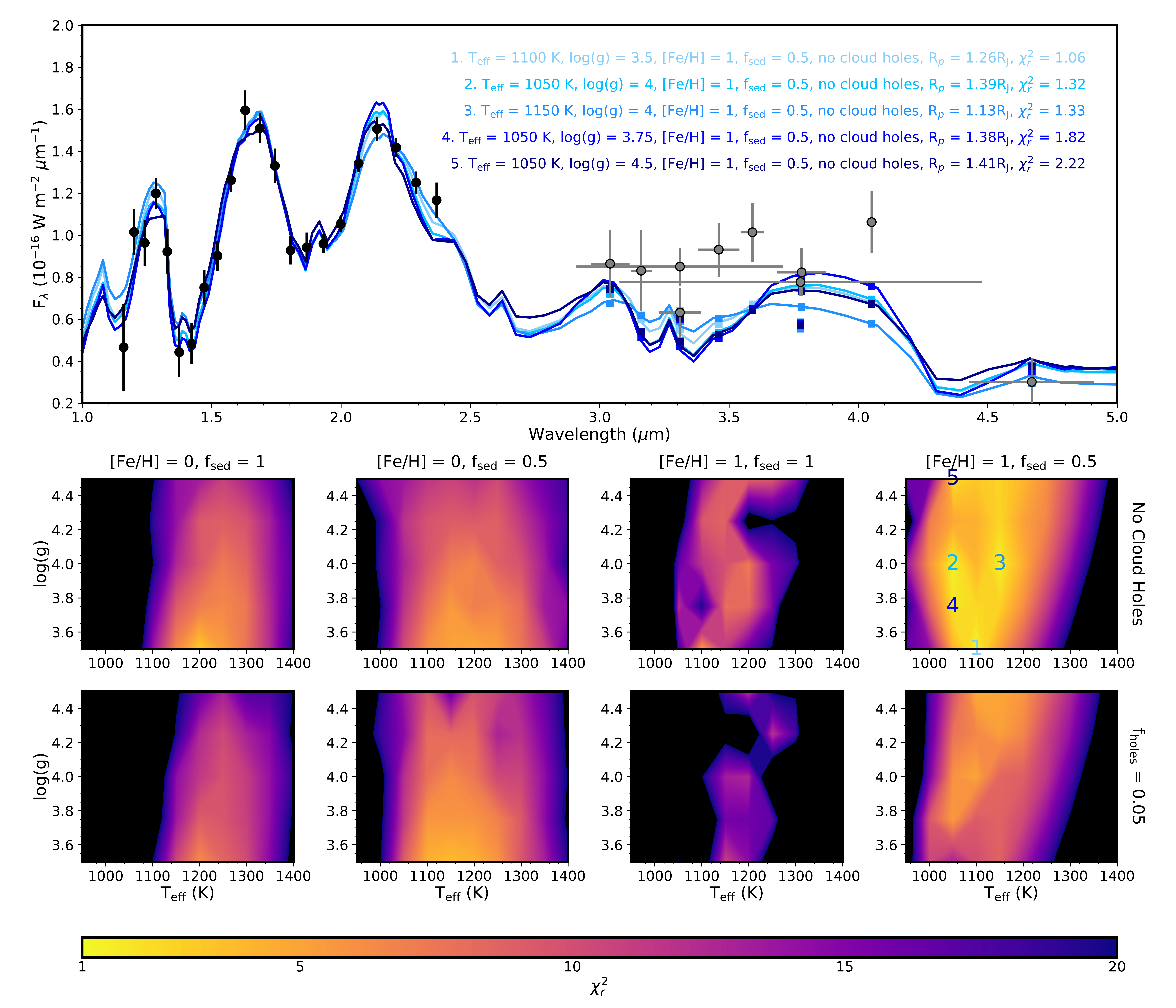}
    \caption{Same as Figure \ref{fig:hr8799c_model}, but for HR 8799~d.}
    \label{fig:hr8799d_model}
\end{figure*}

\begin{figure*}
    \centering
    \includegraphics[width=0.8\textwidth]{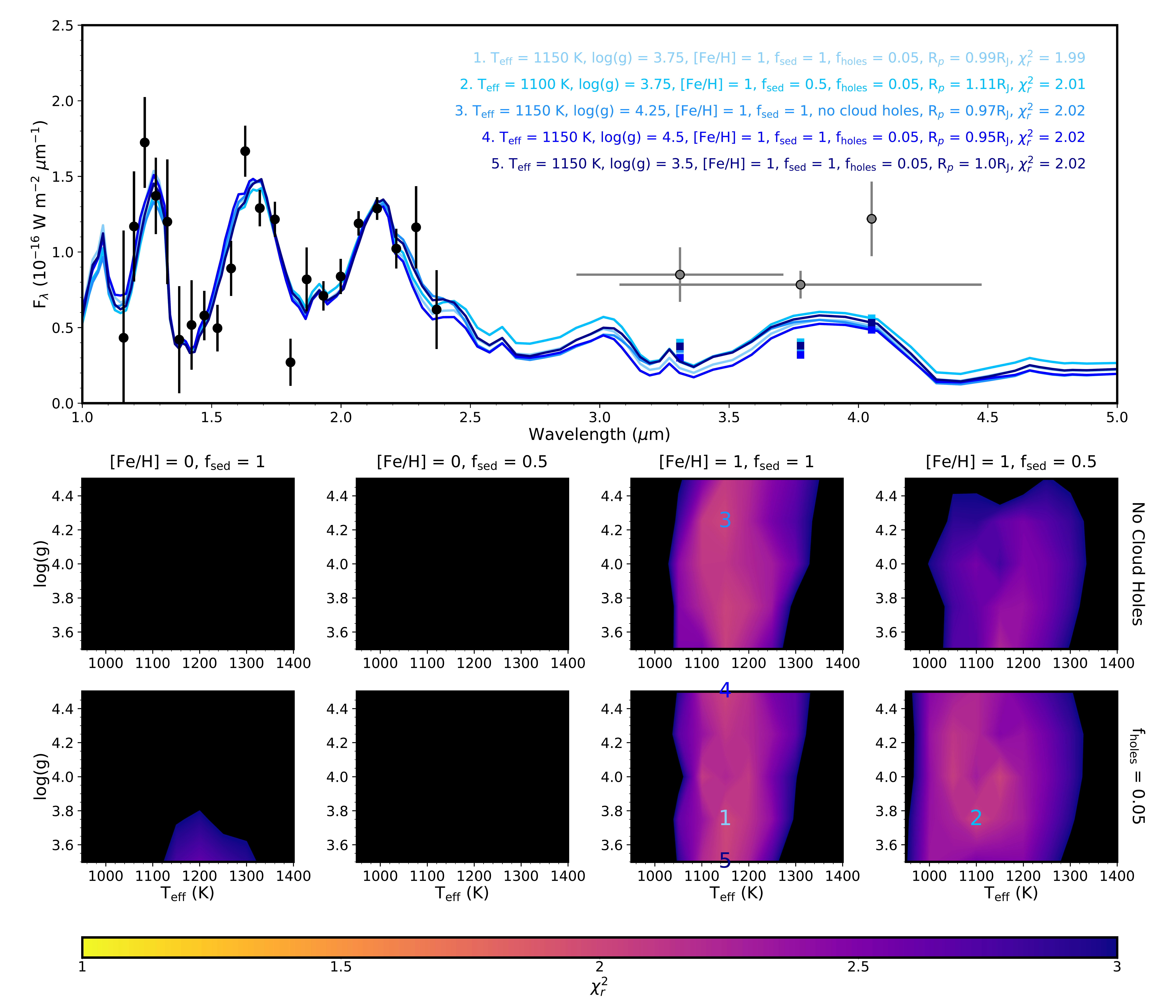}
    \caption{Same as Figure \ref{fig:hr8799c_model}, but for HR 8799~e.}
    \label{fig:hr8799e_model}
\end{figure*}

Even though we did not detect significant atmospheric variability, the long time baseline of our observations allowed us to obtain high SNR spectra with which we can constrain atmospheric properties (Figure \ref{fig:hr8799_spec_fullnight}). These spectra achieve higher SNR than previous spectra from low-resolution integral field spectrographs \citep{Oppenheimer2013, Bonnefoy2016, Greenbaum2018}. Given HR 8799 c, d, and e are similar in luminosity and spectral type \citep[e.g.,][]{Greenbaum2018}, and are locked in mean-motion resonance with similar masses \citep{Fabrycky2010,Wang2018}, identifying and characterizing the similarities and differences in these planets' spectra can be insightful. Atmospheric characterization of the HR 8799 planets have been carried out previously using spectra and photometry on numerous occasions (see Section \ref{sec:intro} and references therein), and therefore we will conduct only a limited investigation here focused on our data set and defer detailed global forward model comparisons and retrieval analyses to future work. 

To aid in our characterization we generate a small custom grid of cloudy radiative-convective-thermochemical equilibrium models using the oft-used thermal structure code EGP \citep[e.g.][]{mckay1989,marley1996,saumon2008,marley2012,Skemer2014,Ingraham2014,Greenbaum2018} with updated opacities from \citet{marley2021}. We assume a solar C/O ratio, as suggested by recent moderate resolution observations \citep{Molliere2020,Ruffio2021}. We vary the effective temperature ($T_\textrm{eff}$) from 950 to 1400 K in steps of 50 K and the log of gravity in cgs units (log(g)) from 3.5 to 4.5 in steps of 0.25; we consider atmospheric metallicities of solar and 10 $\times$ solar; sedimentation efficiencies ($f_\textrm{sed}$) of 1 and 0.5 for silicate and iron clouds; and compare models with uniform clouds and patchy clouds with a fraction of cloud holes ($f_\textrm{hole}$) of 0.05. We base these parameter ranges on previous literature values \citep[e.g.][]{marley2012,Ingraham2014,Bonnefoy2016,Greenbaum2018}. We compare our models to data for each planet by first averaging both nights' data and then optimizing the radius of the model planets to produce a best fit. We compute the reduced $\chi^2$ with 16 degrees of freedom (22 data points minus 6 free parameters: $T_\textrm{eff}$, log(g), atmospheric metallicity, $f_\textrm{sed}$, $f_\textrm{hole}$, and the planet radius). We tested the impact of including literature data at 3-5 $\mu$m in our fits but we found no significant change in our best fit models, only higher $\chi^2$. Systematic shifts in flux calibrations between our data set and previous data sets may complicate joint fits, but when we introduced these shifts as a free parameter we found that the best fit models often tended towards unphysical radii to try to match all the data. We therefore ignored the literature data in our fit, but we do plot them alongside our best fit models to compare them (see below).

We find that HR 8799 c and d are distinct from each other, which can also be clearly seen visually in the spectra (Figure \ref{fig:hr8799_spec_fullnight}). \citet{Greenbaum2018} also found differences between the two planets, although not the same differences: they found a fainter K-band spectra for HR 8799 c which they hypothesize could also be due to photometric calibration offsets. On the other hand, we generally find HR 8799 c to be brighter than HR 8799 d in the $J$, $H$, and $K$ bands, but of comparable brightness in the water absorption bands. While the larger uncertainties on HR 8799 e's spectra result in looser constraints, the shape of the spectrum is more similar to planet c than d (Figures \ref{fig:hr8799c_model}, \ref{fig:hr8799d_model}, \ref{fig:hr8799e_model}). However, its flux level is more comparable to HR 8799 d, so we find that, while HR 8799 e has similar atmospheric parameters as HR 8799 c, HR 8799 e has a smaller radius than HR 8799 c.  

For all three planets we find a preference for supersolar metallicity, $T_\textrm{eff}$'s of $\sim$1100 K, and largely unconstrained log(g) values. Supersolar metallicity has been suggested for all four HR 8799 planets \citep{Barman2011,Bonnefoy2016,lavie2017}, and we confirm this in our analysis. The best fit models for planets c and e do not demonstrate a preference in $f_\textrm{sed}$ or cloud patchiness among our limited parameter search. In contrast, the best fit planet d models consistently favor those with low $f_\textrm{sed}$ (vertically extended) and uniform clouds, suggesting a difference in cloud and/or dynamical processes in the atmosphere of planet d versus planets c and e. A full layer of clouds could be blocking us from seeing deep down into HR 8799 d, causing its $J$, $H$, and $K$ band fluxes to be lower than those of HR 8799 c but keeping its flux in the water absorption band comparable since the water absorption happens above the cloud layer. It should be noted that the model fits to the data for planet c is considerably worse than that for planet d (reduced $\chi^2$ of $>$8 versus $\leq$1.5, respectively); in particular, the best fit planet c model overestimates the J band data by $\sim$30\%, while higher reduced $\chi^2$ models that better fit the J band underestimate the H band flux. In addition, all of our best fit models underestimate the flux between 3 and 4.5 $\mu$m, as compared to literature data taken from \citet{Bonnefoy2016}, by a few to a few tens of percent, while matching the observed $M$ band flux well for planets c and d. Thus, the exact physical process responsible for HR 8799 d being different should be taken with caution, but it may be that HR 8799 d has more extended clouds that cover its entire surface. 

Discrepancies between our models and the data could arise from a number of sources. As our model grid is limited in size and resolution, especially in atmospheric metallicity, $f_\textrm{sed}$, and $f_\textrm{hole}$, we could have simply missed the best fit solution. Furthermore, there are more dimensions to atmospheric composition and clouds that we did not explore here. A more detailed forward model investigation would be needed to solve this problem. 

A major assumption of our work is that of thermochemical equilibrium, which many previous works have already shown to be lacking \citep{Barman2011,marley2012,Skemer2014,Ingraham2014,Molliere2020}. Our models all show a deep methane absorption feature centered at 3.3 $\mu$m, which contributes to the underestimation of the flux between 3 and 4.5 $\mu$m compared to previous observations \citep{Bonnefoy2016}. Disequilibrium processes such as vertical mixing would enrich the photospheric region with CO-rich gas thereby reducing methane absorption. 

The best fit planet radii of our models are $>$1.1R$_J$ for planets c and d and $\sim$1R$_J$ for planet e. Evolution models suggest radii $>$1.1R$_J$ at the ages of the HR 8799 planets, possibly as large as $\sim$1.4R$_J$ \citep{baraffe2003,saumon2008,marley2012}, in agreement with our fits for planets c and d, though we seem to be underestimating the radius of planet e. Underestimation of planet radii by radiative-convective models has been seen previously \citep[e.g.][]{Barman2011,Greenbaum2018} and is likely the result of simplifying assumptions in the model chemistry and physics.

\section{Conclusion}
In this paper, we present two full nights of spectrophotometric monitoring on HR 8799 c, d, and e using the CHARIS spectrograph behind the SCExAO adaptive optics system on Subaru. We first analyzed the use of satellite spots for precise spectrophotometric calibration of the data, as there are no other reference stars in the images. 

\begin{itemize}
    \item The incoherent satellite spots used for photometric calibration have a flux ratio that varies with time with a scatter of at least 3\% and drifts on the 10-minute timescales that can be even larger, so it should be used with caution as a reference for percent-level photometry. 
    \item The spectral (color) stability of the satellite spots is better and experiences much smaller temporal correlations. We find a systematic uncertainty floor of 2\% when measuring the colors of exoplanets using the satellite spots for calibration. 
\end{itemize}

We then extracted the spectra of the HR 8799 planets and performed time series analysis to look for periodic variations. 

\begin{itemize}
    \item We did not observe any significant photometric variability, possibly owing to unfavorable viewing geometry, but{, for rotation periods between $\sim$5-18~hours,} achieved $H$-band sensitivity down to 10\% variability amplitudes for HR 8799 c and 30\% variability amplitudes for HR 8799 d. These are the best constraints yet for these two planets. 
    \item We also did not observe significant variability in the planets' colors, but were sensitive to 20\% variability amplitudes in $H$/$K$ flux ratio for HR 8799 c. 
    \item Our limits exclude the highest variability models for HR 8799 c, and support the theory that these planets are not viewed equator-on, leading to lower variability amplitudes. 
\end{itemize}

Our two nights of monitoring data can be stacked together to produce high signal-to-noise spectra of HR 8799 c, d, and e. We compared the planets' spectra to a small grid of radiative-convective-thermochemical equilibrium models with silicate and iron clouds to characterize their atmospheres.  

\begin{itemize}
    \item From Figure \ref{fig:hr8799_spec_fullnight}, we can visually see that HR 8799 c is brighter than HR 8799 d in the $J$, $H$, and $K$ bands, while HR 8799 e has a similar shape to HR 8799 c, but with lower flux. 
    \item All three planets are best fit with models with $T_\textrm{eff}$ $\sim$ 1100 K and supersolar metallicity, while log(g) is unconstrained within the bounds of our parameter space. 
    \item Our fits do not constrain cloud properties for planets c and e, as both high (1) and low (0.5) $f_\textrm{sed}$ and uniform and patchy (with $f_\textrm{hole}$ = 0.05)  clouds were allowed. In contrast, the best fit models for planet d were consistently those with vertically extended ($f_\textrm{sed}$ = 0.5) and uniform clouds, suggesting that planet d's cloud and/or atmospheric dynamical processes are distinct from those of planets c and e. 
    \item The best fit model planet radii for planets c and d were consistent with evolution models, while that of planet e was underestimated in comparison. 
\end{itemize}

The HR 8799 planets still remain some of the best targets for studying the clouds of exoplanets. Possibly due to an unfavorable viewing geometry, the variability amplitude is not as high as the most photometrically variable planetary mass objects. Future improvements in high-contrast imaging instrumentation can help us reach the necessary precision to detect variability in the light curves of these planets. New ideas to alternate turning the satellite spots on and off to remove the quasi-static speckles at the same location \citep{Sahoo2020} and to generate static, incoherent satellite spots for calibration \citep{Bos2020} are promising avenues to achieve significantly higher photometric calibration precision. Improvements to the SCExAO adaptive optics system \citep{Guyon2021} and new high-contrast imagers coming to Maunakea \citep[e.g. GPI2,][]{Chilcote2020,Marois2020} should provide higher signal-to-noise measurements of the planetary spectra, which is the limiting factor in our analysis. Furthermore, the upgrades to adaptive optics systems should improve signal-to-noise at shorter wavelengths like in the $J$ band where the bulk of substellar object variability has been observed. Obtaining time series data from space using the James Webb Space Telescope would also bypass calibration issues due to observing from the ground. 

\acknowledgments
We thank the anonymous referee for helpful suggestions that improved the manuscript. J.J.W. thanks Jean-Pierre V\'eran for helpful discussions on atmospheric turbulence.
J.J.W. and P.G. were supported by the Heising-Simons Foundation 51 Pegasi b postdoctoral fellowship during the bulk of this research project. 
CHARIS was built at Princeton University under a Grant-in-Aid for Scientific Research on Innovative Areas from MEXT of the Japanese government (\# 23103002)
The development of SCExAO was supported by the Japan Society for the Promotion of Science (Grant-in-Aid for Research \#23340051, \#26220704, \#23103002, \#19H00703 \& \#19H00695), the Astrobiology Center of the National Institutes of Natural Sciences, Japan, the Mt Cuba Foundation and the director's contingency fund at Subaru Telescope. 
The authors wish to recognize and acknowledge the very significant cultural role and reverence that the summit of Maunakea has always had within the indigenous Hawaiian community, and are most fortunate to have the opportunity to conduct observations from this mountain.

\facility{Subaru (SCExAO/CHARIS)}

\software{{\tt pyKLIP} \citep{Wang2015}, \texttt{astropy} \citep{Astropy2013,Astropy2018} }

\clearpage

\bibliography{ref}{}
\bibliographystyle{aasjournal}

\end{CJK*}
\end{document}